\documentclass{aa}  

\usepackage{txfonts}
\usepackage{appendix,epsfig,graphicx,color,mathtools}
\usepackage{arydshln}
\usepackage{placeins}
\usepackage{natbib,twoopt}
\bibpunct{(}{)}{;}{a}{}{,}             %% natbib format for A&A and ApJ
\newcommand{\La}{\mathcal{L}}
\definecolor{Blue}{rgb}{0.06,0.20,0.93}

\def\Hbeta {H$_\beta$}

\def\si31{Si~{\scriptsize III}~$\lambda$4553}
\def\si3{Si~{\scriptsize III}~$\lambda\lambda$4553, 4568, 4575}
\def\si4{Si~{\scriptsize IV}~$\lambda\lambda$4089, 4116}
\def\kms {km~s$^{\rm -1}$\,}

\def\Mspec {$M_{\rm spec}$\,}
\def\Mevol {$M_{\rm evol}$\,}
\def\Minit {$M_{\rm init}$\,}
\def\Teff {$T_{\rm eff}$\,}

\def\vsini {$v \sin i$\,}

\def\logg {$\log g$\,}

\def\vrot{$V_{\rm rot}$\,}

\def\vmac {$v_{\rm mac}$\,}
\def\vmic {$v_{\rm mic}$\,}
\def\vmean {$\overline{v}_{\rm mic}$\,}
\def\vmacmean{$\overline{v_{\rm mac}~(upl)}$\,}

\def\vsinmean{$\overline{\rm vsin i~(upl)}$\,}

\def\Msun {$M_\odot$\,}

\def \ao{\frac{1}{\alpha'}}

\def \beq{\begin{equation}}
\def \eeq{\end{equation}}
\def \ben{\begin{enumerate}}
\def \een{\end{enumerate}}
\def \beqa{\begin{eqnarray}}
\def \eeqa{\end{eqnarray}}

\begin{document}

   \title{Micro-turbulence across the Hertzsprung-Russell diagram. Observationa constrains 
   for stars in the MW} 

\author{N. Markova\inst{1}, M. Cantiello\inst{2,3} \and L. Grassitelli
}

\institute{Institute of Astronomy, National Astronomical Observatory,
  Bulgarian Academy of Sciences, P.O. Box 136, 4700 Smolyan, Bulgaria
 (\email{nmarkova@astro.bas.bg})
 \and
Center for Computational Astrophysics, Flatiron Institute, 162 5th
Avenue, New York, NY 10010, USA
\and
Department of Astrophysical Sciences, Princeton University, Princeton,
NJ 08544, USA
}

\abstract{Despite its critical importance for determining stellar properties and evolution, the origin and physical nature of micro-turbulence remains poorly understood. Most of the existing works are focused on specific spectral types and luminosity classes, and a comprehensive,  unified view has yet to emerge.}
{ Our main goal is to investigate the behaviour of photospheric micro-turbulence  across the Hertzsprung–Russell (HR) diagram and to bridge theory with observations.}
{ We assemble a homogeneous database of precise and consistent determinations of effective temperature, surface gravity, projected rotational rate (\vsini) and  macro- and micro-turbulent velocities (\vmac \& \vmic) for over 1800 Galactic stars spanning spectral types O to K and luminosity classes I to V. By carefully minimizing biases due to target selection, data quality, and disparate analysis techniques, we perform statistical tests  and comparative analyses to probe potential dependencies between these parameters and \vmic.}
{ Our findings indicate that photospheric micro-turbulence is a genuine physical phenomenon rather than a modelling artifact. A direct comparison between observed \vmic velocities  and corresponding theoretical predictions for the turbulent pressure fraction strongly suggests that this phenomenon most likely arises from photospheric motions driven (directly or indirectly) by envelope convection zones, with an additional pulsational component likely operating in main-sequence B stars. We show that neglecting micro-turbulence in Fourier transform  analyses can  partly  (but not solely) explain the dearth of slow rotators and the scarcity of stars with extremely low \vmac. We argue that including micro-turbulent pressure in atmospheric modelling can significantly mitigate (even resolve) the mass discrepancy for less massive O stars.}
{ We provide new observational insights into the nature and origin of micro-turbulence across the HR diagram. Our database offers a valuable resource for testing and refining theoretical scenarios, particularly those addressing puzzling phenomena in hot massive stars.}
\date{Received <>/Accepted <27.07.2025>}

\titlerunning{Micro-turbulent broadening  of Galactic stars}
\authorrunning{Markova et al. }
\maketitle

\section{Introduction}\label{intro}

Micro-turbulence (\vmic) is a key parameter in one-dimensional (1D) model atmosphere analyses of stellar spectra. Due to its influence on non-Local Thermodynamic Equilibrium  (NLTE) occupation numbers (and hence the atmospheric structure) and on formal integral calculations--primarily of metal lines but also of He I \citep{SH98, McE99, massey13, markova20}--this quantity is frequently introduced as a free parameter, particularly in the O-star regime, to achieve better agreement between model predictions and observations. However, this practice can introduce uncertainties in empirically derived stellar properties, carrying significant implications for our understanding of stellar physics and evolution \citep[e.g.,][]{gonzalez12, markova14, markova18, markova20}.

Despite some unresolved issues \citep{mucciarelli11}, the prevailing interpretation of micro-turbulence is that it represents non-thermal velocity fields in stellar atmospheres at scales shorter than the photon mean free path. Consequently, in hot massive stars whose atmospheres are dominated by radiation pressure, micro-turbulent broadening has often been considered an artifact resulting from imperfect modelling of physical processes in stellar atmospheres. These are related to macroscopic velocity fields within extended atmospheres \citep{kud92}, deviations from local thermodynamic equilibrium (LTE) in line formation calculations used as diagnostics for \vmic \citep{prz01, prz06}, stellar pulsations \citep{townsend07}, or the presence of subsurface convection zones \citep{cantiello09, grassitelli15a, grassitelli15b}.

Conversely, in low-mass stars with atmospheres that are partially or predominantly convective, the physical reality of photospheric micro-turbulence is difficult to dismiss. Empirical \citep{edmunds78} and theoretical \citep{asp20a, asp20b} evidence suggests that in these stars, micro-turbulent broadening is primarily related to convective processes that are not accounted for in standard 1D model atmosphere calculations.

Thus, despite its crucial role in determining stellar properties and evolution, the origin and physical nature of micro-turbulence remain poorly understood. Existing studies have typically concentrated on specific spectral types (SpT) and luminosity classes (LC), lacking a unified global perspective. To address this gap, we initiated a project aimed at providing a comprehensive and statistically significant overview of the empirical properties of photospheric micro-turbulence across a diverse range of Galactic objects. The main outcomes of this project are presented and discussed in this paper.

\section{Data Collection}\label{general}
\subsection{Issues and strategy}\label{strategy}

An important consideration when compiling and merging multiple data-sets from various literature sources is that these sources can differ significantly in terms of target selection criteria, data quality, and the methodologies employed to determine stellar properties. Consequently, any consolidated dataset of literature \Teff, \logg, and \vmic\ values necessary for our analysis will inherently be susceptible to various biases. Some of these biases can affect the precision of our analysis, while others could impact the overall accuracy and reliability of our final results.

To ensure a certain degree of consistency among the datasets collected from the literature, we developed and implemented the following strategy:
\ben

\item[1)]{To minimize potential biases and simplify our analysis, we restricted our study to (presumably) single pulsating and non-pulsating stars spanning various evolutionary stages  from the main sequence (MS) to the red giant/supergiant phases. \footnote{Wolf-Rayet stars and Luminous Blue Variables were excluded due to their complex physics and lack of reliable \vmic\ diagnostics.} We aimed to maintain a balanced representation across various SpT and LC, thereby achieving comprehensive coverage of the spectroscopic Hertzsprung-Russell diagram (sHR diagram; \citealt{LK14}). \footnote{An analogue of the classical HR diagram where stellar luminosity is replaced by the quantity $\La/\La_{\odot}$ = \Teff$^4$/$g$, effectively representing the luminosity-to-mass ratio.}}

\item[2)] {To address observational biases, we exclusively considered studies utilizing  high-resolution photometric and spectroscopic observations with high signal-to-noise ratios.}

\item[3)]{To mitigate methodological discrepancies, we prioritized \Teff, \logg, and \vmic\ values derived from proven methods known to deliver accurate and reliable results. We specifically selected methodologies that are mutually consistent and show  minimal systematic discrepancies between them. \footnote{Systematic uncertainties here refer to those arising from different modelling approaches employed to determine \Teff, \logg, and \vmic.}}
\een

While the first two criteria were relatively straightforward to meet, the third criterion proved more challenging because: i) \Teff\ and \logg\ for most stars are not directly measurable and depend on model-dependent methods, and ii) stellar physics strongly depends on initial mass (\Minit) and evolutionary stage, necessitating various specialized model atmospheres and line-formation codes tailored for specific mass and temperature regimes, as well as wavelength ranges. 

To overcome this issue, we conducted an extensive review of the literature to identify state-of-the-art model atmospheres, line-formation codes, and analytical techniques frequently employed for specific stellar types, and thus considered credible.  Although three-dimensional radiative hydrodynamic model atmospheres and spectra are available (e.g., \citealt{husser13}), we have opted to use one-dimensional models exclusively for consistency. Similarly, for consistency, we limit our analysis to optical (or combined UV/optical) data (see the next section).

\subsection{Model atmosphere and line formation codes }\label{models}

From the extensive array of model atmosphere and line-formation tools available for determining stellar properties (see \citealt{sander17} for hot massive stars and \citealt{jofre19} for low-mass stars), we selected the methods listed in Table~\ref{B1} as sufficiently robust for populating our database.

Specifically, for massive stars with strong winds (OB and early A-type supergiants), we relied exclusively on data generated by FASTWIND (FW) and CMFGEN, both of which consistently account for UV line-blocking/blanketing effects and significant departures from Local Thermodynamic Equilibrium (LTE). For stars with weaker or negligible winds, we preferred a combined approach involving non-LTE line formation calculations (SYNSPEC or DETAIL/SURFACE) built upon NLTE (TLUSTY) or LTE (ATLAS9) hydrostatic, plane-parallel, line-blanketed model atmospheres. This approach has demonstrated reliability comparable to purely NLTE analyses \citep{dufton05, prz06, NP07}, and was our primary choice for late-O and B-type dwarfs and less luminous AF supergiants.

Finally, for low-mass stars (giants and supergiants with \logg$<\,$2 dex apart; see \citealt{HE06}), whose physical conditions are generally well-represented by LTE, hydrostatic equilibrium, and plane-parallel geometry, we primarily utilized the LTE codes ATLAS9 and MARCS combined with various LTE radiative transfer tools. These methodologies were considered reliable sources for our investigation (see Sect.~\ref{accur_spec} for further details).

\subsection{Analysis techniques to determine stellar temperatures and
gravities}\label{sq_technics}

To ensure a meaningful \vmic\ analysis, the literature datasets employed must be not only accurate and reliable but also precise. Since precision depends not only on observational quality but also on the analytical techniques used, we selected six of the most widely adopted methods from the literature (for a comprehensive review, see \citealt{smalley05}). These methods, considered sufficiently precise and trustworthy within their respective ranges of applicability, include three techniques employed in spectroscopic analyses and three used in photometric analyses.

\subsubsection{Spectroscopic techniques}

The two most commonly used spectroscopic techniques for determining \Teff\ and \logg\ in various stellar types (excluding classical Cepheids, see below) are the global line profile fitting (LPF) approach and the method based on fitting equivalent widths (EWs). Both methods are considered in our analysis, with the LPF approach preferred for high-mass stars, where both line strength and profile shape provide essential diagnostics, and the EW method predominantly employed for lower-mass stars (column 5 in Table~\ref{spdat}).

For classical Cepheids—whose temperatures and surface gravities vary significantly during their pulsation cycles—a specialized approach combining empirical calibrations of line-depth ratios (LDRs) for determining \Teff, and a comparison between observed and synthetic EWs for estimating \logg\ and \vmic, has proven more suitable. This combined approach, hereafter referred to as the "LDRs/EW approach", is the method we adopt for collecting data necessary to analyse the \vmic\ properties of classical Cepheids.

\subsubsection{Photometric techniques}

Although the optimal method to determine photospheric parameters for a given type of star is through consistent spectroscopy and methodology, accurate and precise values for \Teff, \logg, and \vmic\ can also be reliably derived from photometric observations. To enhance the statistical robustness of our analysis, particularly for low-mass stars, we selected three photometric techniques considered sufficiently accurate and reliable for inclusion in our database (see Table~\ref{photdat}). Specific information on these methods and their applicability ranges are summarized below; additional details can be found in \citet{smalley05}.

\textbullet{[M1] \Teff\ and \logg\ derived from comparisons between predicted and observed reddening-free Johnson $Q$ and/or Strömgren [c1] and $\beta$ indices.
\footnote{$Q = (U-B) - 0.72(B - V)$; $[c1] = c1 - 0.20(b - y)$; $\beta$ measures the intensity (equivalent width) of the \Hbeta\ line.}
Due to the negligible Balmer discontinuity and nearly degenerate UBV intrinsic colours in hot massive stars, this technique is reliable primarily for mid-to-late B-type dwarfs and cooler stars.
}

\textbullet{[M2] \Teff, \logg, and \vmic\ determined by fitting synthetic fluxes to de-reddened spectral energy distributions (SEDs) combined with distance measurements from Hipparchus/Gaia parallaxes. This method is effective only for stars whose SEDs  is not affected by wind effects, and thus, can be accurately modelled using five parameters (\Teff, stellar radius (R), \vmic, metallicity, and interstellar reddening). \logg\ is constrained from the derived temperature and radius \citep{gray01}.
}

\textbullet{[M3] \Teff\ obtained from calibrations of photometric colours; \logg\ derived either by fitting synthetic to observed Balmer/metal lines [M3a] or from absolute bolometric magnitudes [M3b]. Since available photometric calibrations generally rely on flux distributions from LTE, line-blanketed Kurucz (1979, 1991) model atmospheres
\footnote{These calibrations typically do not account for gravity-dependent effects in the spectral energy distribution, which become significant in extended atmospheres of hot, massive stars.},
this method is primarily applicable to low-mass stars \citep{luck18a, ivanyuk17}, massive stars with weak or negligible winds \citep{daflon99, daflon01, smartt02}, and red giants \citep{wang17}.
}

\subsection{Micro-turbulent velocity determinations}\label{vmic_data}

Although very precise and unrestricted in terms of \Teff\ and \logg, the classical method for determining photospheric micro-turbulence —- eliminating trends between derived abundances and equivalent widths of metal lines from a particular ion—is highly model-dependent. Consequently, this approach can lead to substantial systematic differences between estimates derived using LTE versus NLTE analyses \citep{kilian94a, vranken00, prz06}.
\footnote{Significant errors can also emerge from variations in the number of lines used and the balance between weak and strong transitions; however, these errors are challenging to quantify and control.}

To mitigate this issue, we adopted a twofold strategy: For hot, massive stars, where NLTE effects are critical, we selected studies utilizing NLTE line formation calculations (at least for the key diagnostic lines) combined with Si~III lines as a main \vmic\ indicator
\footnote{For early B-stars, O~II lines can also serve as diagnostics; however, see the remarks by \citealt{hunter07}.};
for cooler and lower-mass stars, we relied on scientific studies using LTE calculations paired with \vmic\ diagnostics that are less sensitive to NLTE effects (typically lines of Fe I, but also of Fe~II, Mg~II, and Ti~II; see, e.g., \citealt{SP08, prz01, luck18a, luck18b, ivanyuk17}). Details are summarized in Columns~6 and 3 of Tables~\ref{photdat} and \ref{spdat}).

\section{The sample and the database}\label{data}

Following the strategy detailed in the previous section, we have compiled a comprehensive database containing \Teff, \logg, and \vmic\ determinations for over 1800 presumably single Galactic stars, spanning various spectroscopic characteristics and pulsational properties.
\footnote{We acknowledge that some objects might overlap across different datasets.}
An overview of these stars categorized by spectral type  and luminosity class, along with specific methodological details and references to the corresponding analyses, are presented in Tables~\ref{photdat} and \ref{spdat}. To enhance statistical coverage, we have also included \Teff, \logg, and \vmic\ determinations from \citet{kilian91, kilian94a, kilian94b} and \citet{lehmann11} for a limited number of late-O to early-A type dwarfs and giants, even though these studies employed non-standard model atmospheres and line-formation codes. Additionally, to maintain balanced representation across various SpT and LC, not all stars analysed in each study were necessarily included in our database (see Section~\ref{strategy}).

\subsection{Completeness and biases}\label{distr_vmic}

\begin{figure}
\resizebox {\hsize}{!} {\includegraphics{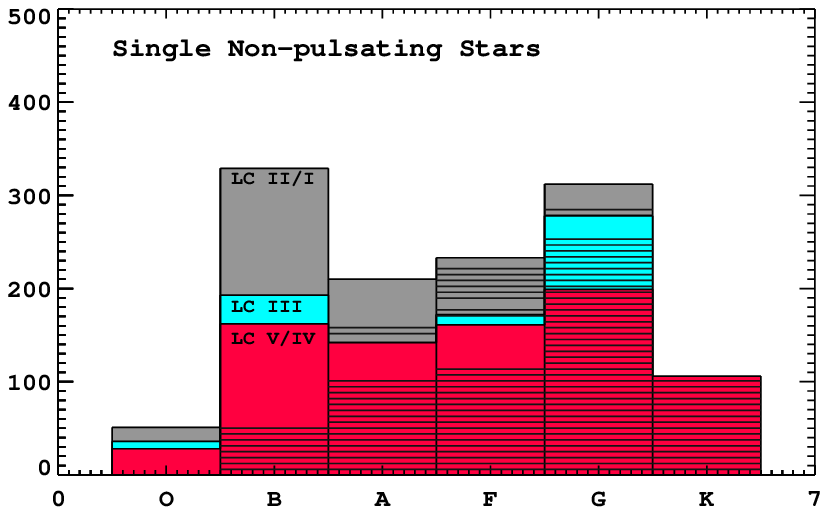}}\\
\resizebox {\hsize}{!}{\includegraphics{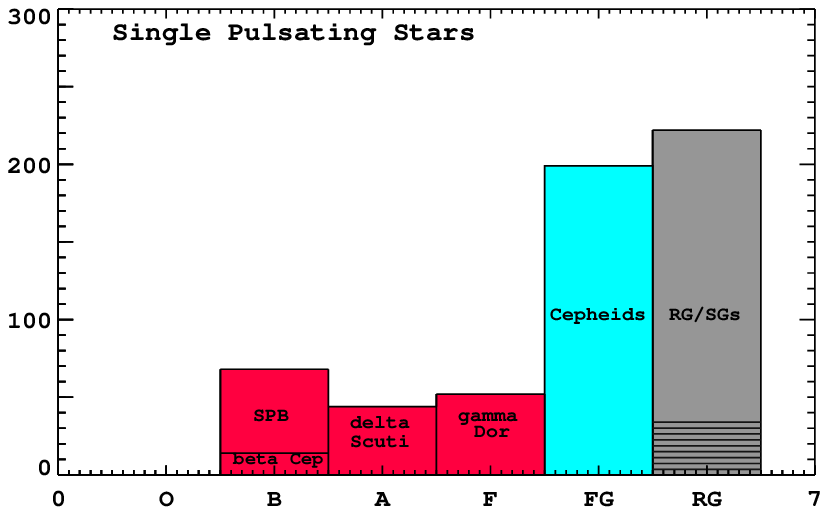}}
\caption{Distribution of the sample stars by spectral type  and luminosity class (LC~V/IV -- red; LC III -- light blue;  LC II/I -- grey), and by 
the type of pulsations. Upper panel -- non pulsating stars; lower panel -- 
sample pulsators. In both cases shaded areas indicate the number of objects with photometric \Teff\ and \logg\ determinations.
}
\label{fig1}
\end{figure}

The distribution of stars in our sample by spectral type and luminosity class  is illustrated in Fig.~\ref{fig1}, with shaded regions indicating the number of objects for which photometric \Teff\ and \logg\ determinations are available. Non-pulsating and pulsating stars are displayed separately (upper and lower panels, respectively). Pulsating stars constitute approximately 20\% of the total sample and include four categories of non-radial pulsators—Slowly Pulsating B-stars (SPBs), $\beta$ Cephei, $\gamma$ Doradus, and $\delta$ Scuti variables—as well as one category of radial pulsators (classical Cepheids). Additionally, a substantial number of red giants/supergiants (RGs/SGs), representing the advanced stages of stellar evolution toward the red portion of the Hertzsprung-Russell diagram (HR diagram), have been included for completeness.

Fig.~\ref{fig1} demonstrates that, overall, the balance among various luminosity classes is relatively good, with pulsating and non-pulsating LC V/IV stars (highlighted in red) comprising nearly half of the entire sample. However, the distribution by spectral type is more heterogeneous: O-type stars appear under-represented, A-type stars of LC III are nearly absent, and FGK-type stars (including classical Cepheids and RGs/SGs) dominate, constituting about 63\% of the total sample.

\begin{figure*}
\begin{center}
\resizebox {\hsize}{!}{\includegraphics{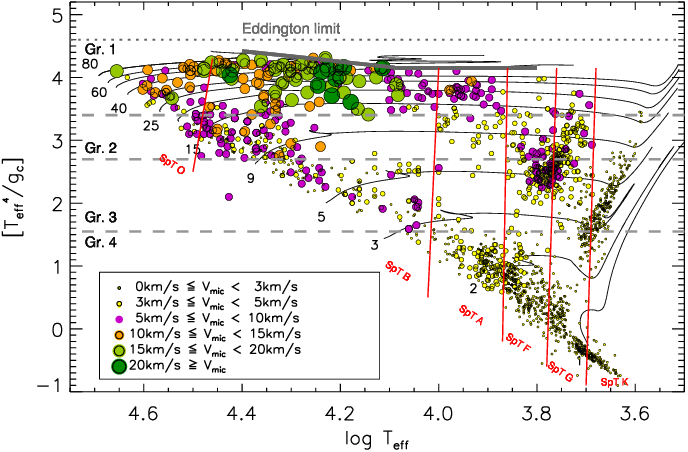}}
\caption{Spectroscopic HR diagram of  the sample stars with data-points colour-coded
and size-scaled according to their \vmic\ value, as indicated in the legend.
Overplotted are: the \citet{brott11} evolutionary tracks for single stars with solar metallicity and initial rotational velocity \vrot = 300~\kms;  the  Eddington limit (horizontal dotted line); the observed HD limit (gray solid thick lines) and  the temperature boundaries corresponding to each spectral type (red almost vertical lines).
Error bars are omitted for clarity. For more information see Sect.~\ref{vmic_shrd}}
\label{fig2}
\end{center}
\end{figure*}

These observations might suggest that our observational dataset could be under-representative in the regime of more massive stars. However, this concern is mitigated by a visual comparison of the sHR diagram in Fig.~\ref{fig2} with similar diagrams from \citet{castro14} and \citet{holgado20}. Such comparisons clearly show that our diagram qualitatively matches those from the cited studies, accurately capturing essential features such as the main and terminal-age main sequences, the Cepheid instability strip, the observed upper boundary of $\La/\La_{\odot}$, and indications of the Humphreys-Davidson limit \citep{HD79}. One notable exception is the somewhat under-represented red supergiant (RSG) branch. Moreover, all diagrams consistently reveal a sparsely populated region in the top-left corner, where the most massive and unevolved O-stars typically reside. While this characteristic is well documented \citep{markova14, massey16, holgado20, SS20, castro21}, it appears more pronounced in our dataset due to the limited availability of reliable \vmic\ diagnostics in optical spectra for O-stars.

The observed deficiency of LC~III A-type stars in our sample is also readily explainable. Their location on the sHR diagram corresponds to the well-known Hertzsprung gap, an evolutionary stage characterized by rapid stellar evolution, resulting in very few observable stars within this region.

\subsection{Accuracy and internal consistency}\label{accur_spec}

From Column 7 of Table~\ref{spdat}, it appears that the accuracy of the
spectroscopic \Teff, \logg\ and \vmic\ determination  in our database
is reasonably good  with typical error for low mass stars somewhat smaller
(higher accuracy) than for the high mass ones: $\Delta$\Teff\  between
$\pm$0.05 and  $\pm$0.10~kK vs. $\Delta$\Teff\ between $\pm$0.2 and
$\pm$1.0~kK; $\Delta$\logg\ between $\pm$0.1 and $\pm$0.3~dex, and
$\Delta$\vmic\ between $\pm$0.04 and $\pm$0.50~\kms vs. $\Delta$\vmic\
between $\pm$1.0~\kms and $\pm$5.0~\kms. The accuracy of the photometric
\Teff\ and \logg\ is also good with 0.08~kK$\lesssim\Delta$\Teff$\lesssim$1.2~kK
and 0.1~dex$\lesssim0.1\Delta$\logg$\lesssim$ 0.3~dex (see Column~6 of
Table~\ref{photdat}).

While these results are most likely a consequence of our data selection strategy
(Sect.~\ref{general}),  one must bare in mind that in some cases (bold-faced
numbers in Tables~\ref{photdat}  and \ref{spdat}) the error provided by the
corresponding authors represents the precision rather than the total accuracy
of the obtained estimates, and must be therefore considered as an upper limit.

Regarding the internal consistency of the adopted datasets, controlling this factor is challenging due to the variety of literature sources involved. Nevertheless, substantial theoretical and observational evidence supports the following conclusions:

\begin{itemize}
\item[1.] The internal consistency among the methodologies and datasets underpinning our analysis is generally robust, with discrepancies typically on the order of the associated uncertainties (see, for instance, \citealt{P05, martins05b, gonzalez12, massey13, Mcevoy15, holgado18, markova20} for massive stars, and \citealt{daflon01, gustaffson08, gebran10, lyubimkov15, ivanyuk17} for lower-mass stars).

\item[2.] The overall agreement between photometric and spectroscopic methods employed in our study is also favourable, with systematic offsets comparable to or only slightly exceeding the associated internal uncertainties \citep{palacio16, cordero14, niemczura15, luck18a}.
\end{itemize}

In summary, although our database is not statistically complete, it is sufficiently extensive to facilitate a meaningful and realistic analysis of the global \vmic\ properties of Galactic stars.

\section{Micro-turbulence as a function of effective temperature and equatorial
surface gravity}\label{vmic_spt}

The \vmic\ properties of the sample stars were investigated by categorizing them according to their spectral type (SpT), luminosity class (LC), and type of pulsation. For non-pulsating stars of a given SpT, the Spearman rank correlation 
test was consistently applied to evaluate the strength ($\rho$) and significance ($p$) of potential relationships 
between \vmic\ and both \Teff\ and \logg
\footnote{In a few cases, this analysis was complicated by substantial scatter in the data or by a limited number of stars with reliably determined stellar properties.}.
Further details are provided in Sect.~\ref{non_puls}. For the pulsating ones, a somewhat different approach has been applied instead (Section~ \ref{vmic_puls} and Appendix~\ref{A}).
\begin{figure*}
\begin{center}
{\includegraphics[width=7.23cm,height=3.71cm]{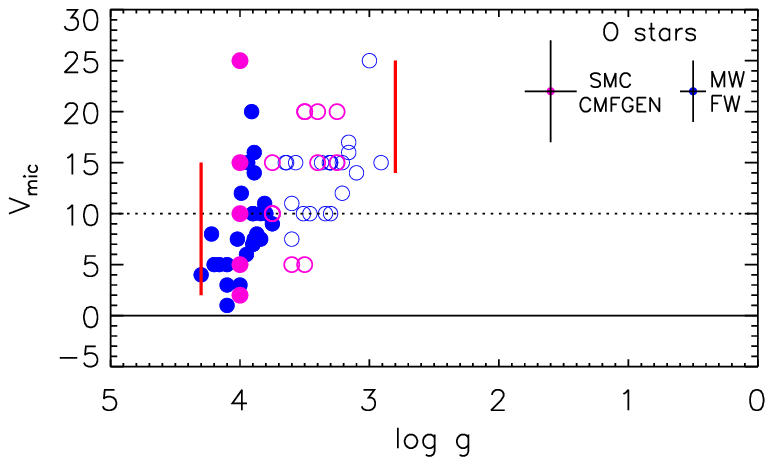}}
{\includegraphics[width=7.23cm,height=3.71cm]{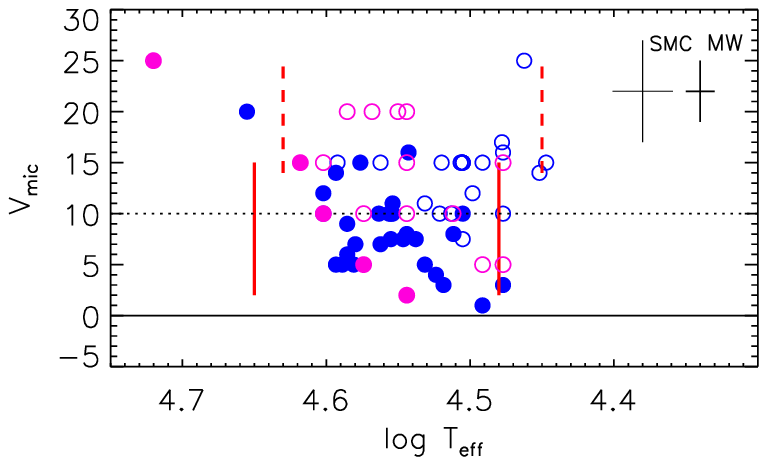}}\\
{\includegraphics[width=7.23cm,height=3.71cm]{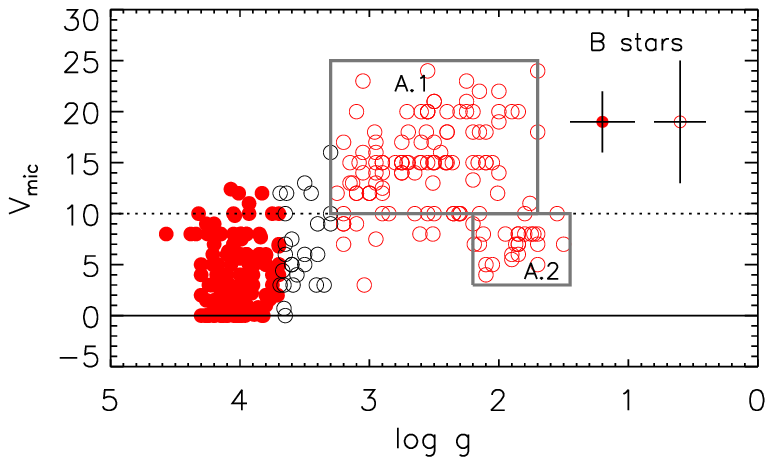}}
{\includegraphics[width=7.23cm,height=3.71cm]{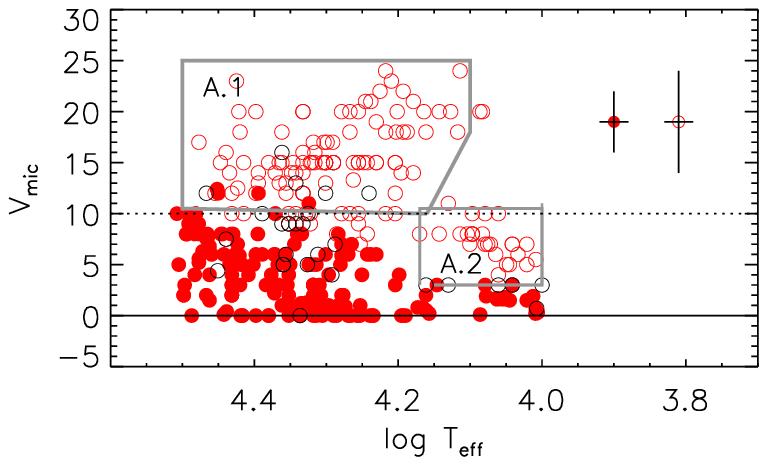}}\\
{\includegraphics[width=7.23cm,height=3.71cm]{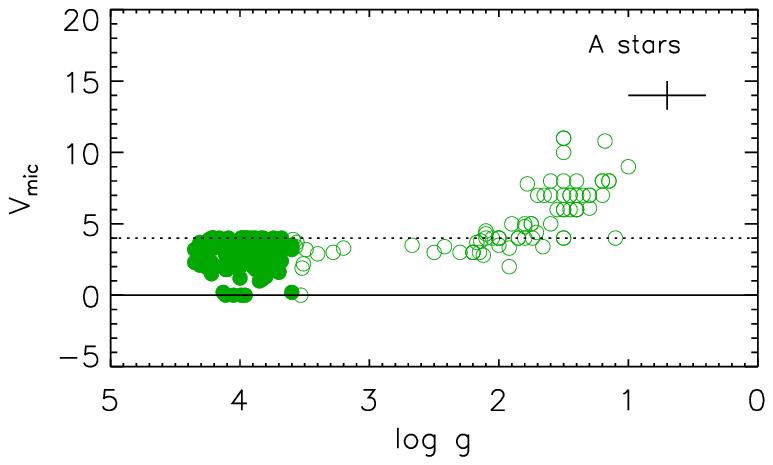}}
{\includegraphics[width=7.23cm,height=3.71cm]{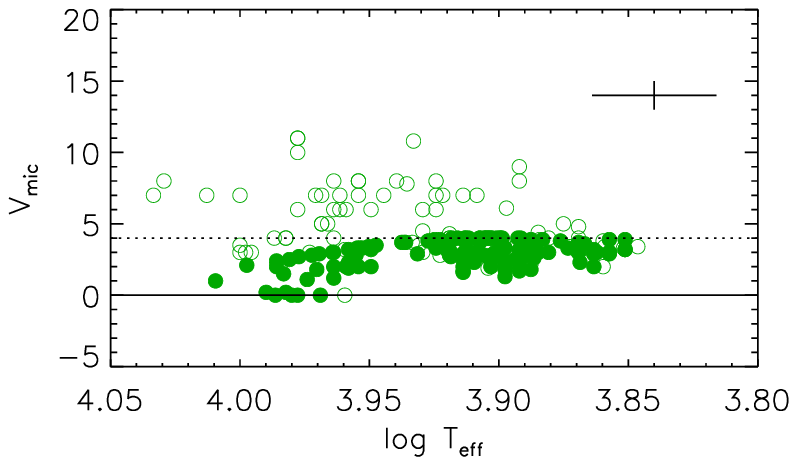}}\\
{\includegraphics[width=7.23cm,height=3.71cm]{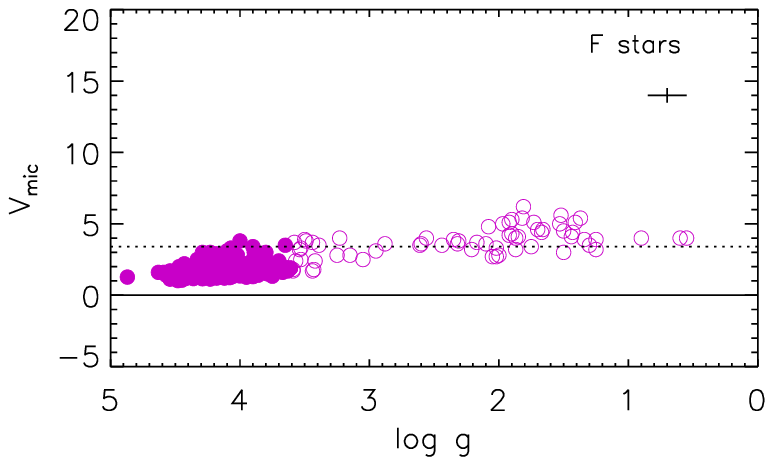}}
{\includegraphics[width=7.23cm,height=3.71cm]{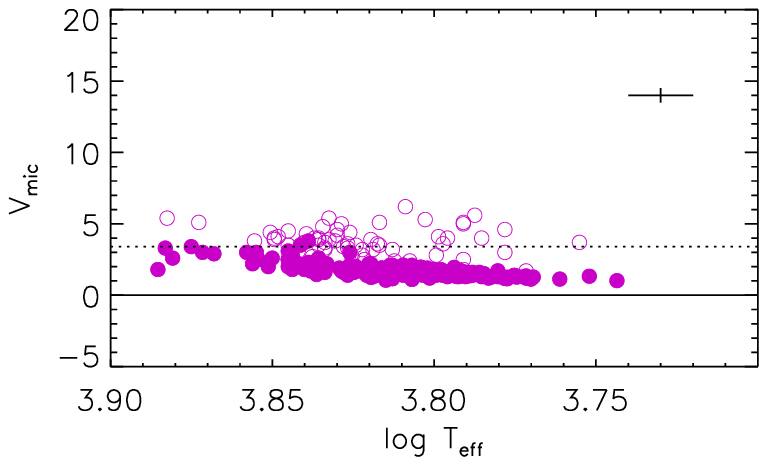}}\\
{\includegraphics[width=7.23cm,height=3.71cm]{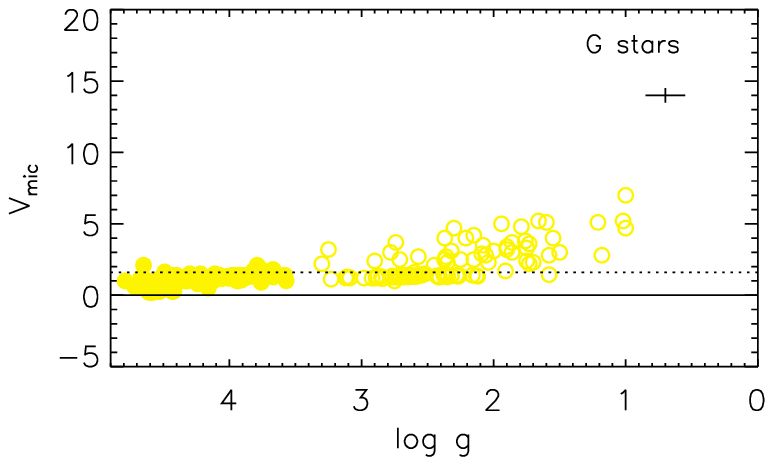}}
{\includegraphics[width=7.23cm,height=3.71cm]{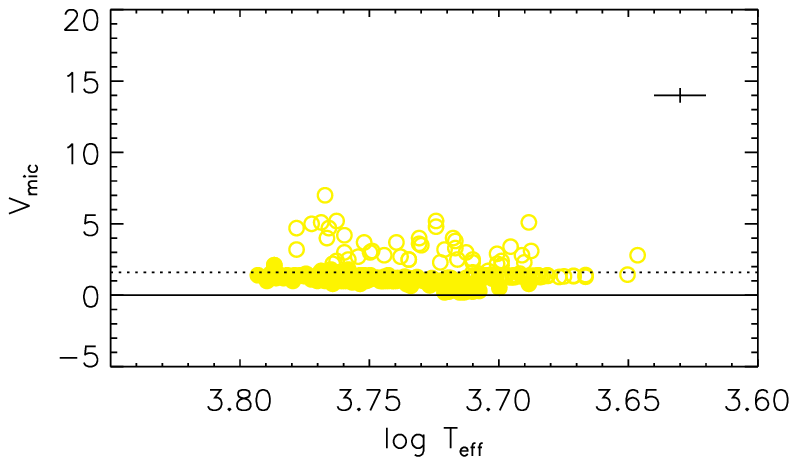}}\\
{\includegraphics[width=7.23cm,height=3.71cm]{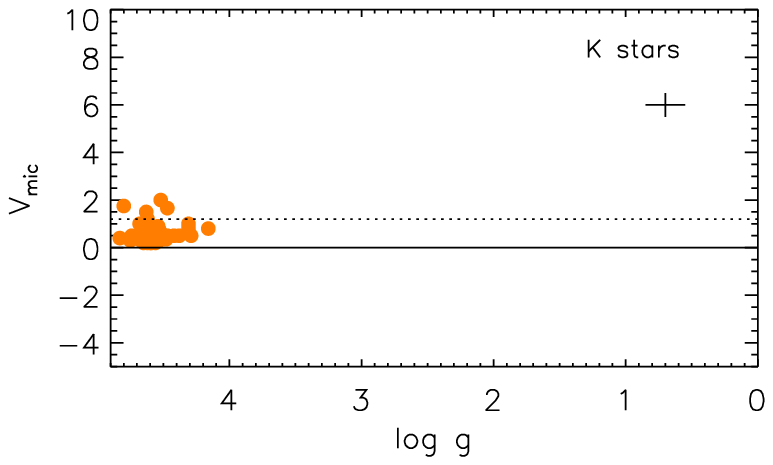}}
{\includegraphics[width=7.23cm,height=3.71cm]{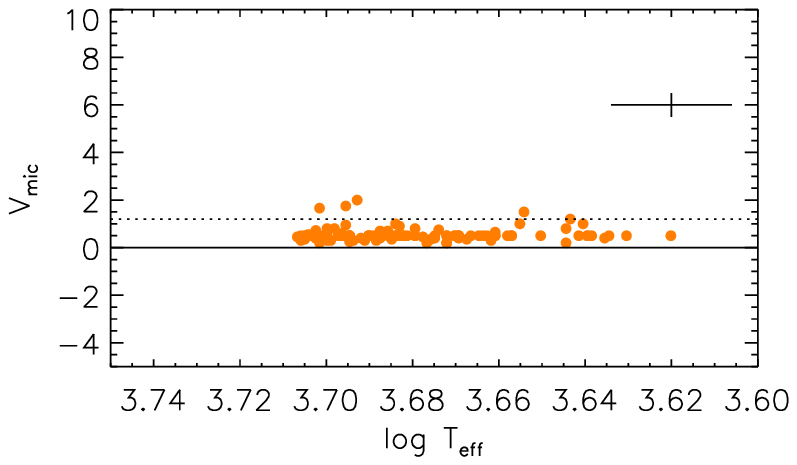}}
\caption{Photospheric micro-turbulence of  the sample non-pulsating stars separated by SpT and LC (open circles - LC~III/I; solid circles - LC V/IV) as a function of \logg\ (left) and \Teff (right). In the top panels over-plotted  are: the \vmic velocities of a sample of O stars in the SMC derived by means of the CMFGEN code (from \citealt{heap06}), and the theoretical \Teff\ and \logg\ scales of O stars in the MW predicted by \citet{martins05a}(red solid (DWs) and dashed (SGs) vertical lines).  In each plot representative error bars are also indicated. For more information, see Sect.\ref{vmic_spt}.
}
\end{center}
\label{fig3}
\end{figure*}

\subsection{Non-pulsating stars: results by spectral type and luminosity
class}\label{non_puls}

In chemical abundance studies, spectral type (SpT) and luminosity class (LC) are frequently—if not always—used as reference points to describe and interpret the \vmic\ properties of the analysed stars. To facilitate direct comparison with such studies, we began our analysis by examining the behaviour of \vmic\ in non-pulsating sample stars as a function of \Teff\ and \logg, separating them by SpT and LC. The key outcomes of this analysis are presented in Fig.\ref{fig3} and summarized in Table\ref{stat_prop}. Detailed comments are provided below.

\paragraph{O-type stars}\label{spt_O}
As discussed in Sect.\ref{distr_vmic}, the number of O-type stars in our database is relatively small: 50 stars in total, including 29 (~58\%) of LCV/IV, 15 (~30\%) of LCII/I, and 6 (~12\%) of LCIII.
\footnote{For stars with \Minit$\ge$ 40 \Msun, this LC classification is admittedly a simplification.}
For most of these stars (37), the relevant data were compiled from the literature. For the remaining stars, \Teff\ and \logg\ values were taken from \citet{markova18}, and \vmic\ was determined by us using He~I~6678 as a diagnostic line (see \citealt{holgado18}, \citealt{markova20}, and references therein).

\begin{table*}[t]
\footnotesize
\begin{center}
\caption[]{Statistical properties of photospheric micro-turbulence of the sample
non-pulsating stars separated by SpT and LC.   }\label{stat_prop}
\tabcolsep1.9mm
\begin{tabular}{llrclcllr}
\hline
\hline
SpT&LC$^{a}$&\# of&log\Teff&\logg&\vmic limits &$\rho$, $p$ &$\rho$, $p$  &\vmean  \\
 &  &stars &coverage$^{b}$ &coverage &[\kms]& (\vmic, \logg) &  (\vmic, log\Teff)  &[\kms]\\
\hline
\\
O  &all stars &50  &4.65 -- 4.45 &4.30 -- 3.20 &1.0 -- 20.0 &{\bf $-$0.74, 9.59$^{-10}$}&$-$0.26, 0.06 &10.6$\pm$4.6\\
   &V/IV      &29  &    ....     &4.30 -- 3.70 &1.0 -- 15.0 &{\bf $-$0.59, 0.00} &$+$0.33, 0.07    &8.0$\pm$4.1\\
   &III/I    &21  &    ....     &3.69 -- 3.20 &5.0 -- 20.0 &{\bf $-$0.44, 0.04} &$-$0.30, 0.18 &14.3$\pm$    2.8\\
\hdashline
B  &all stars &323 &4.50 -- 4.00 &4.40 -- 1.50 &0.0 -- 24.0 & .... & ....& 8.2$\pm$6.2\\
   &V/IV      &158 &   ....      &4.40 -- 3.70 &0.0 -- 12.4   &$-$0.15, 0.06     & {\bf $+$0.48, 1.04$^{-10}$} & 3.8$\pm$3.2\\
   &III       & 28 &    ....     &3.69 -- 3.30 &0.0 -- 16.0  &$-$0.35, 0.07   &{\bf $+$0.50, 0.00}    &6.9$\pm$4.0
\\
   &II/I      &137 &    ....     &3.29 -- 1.50 &3.0 -- 24.0  &    ....       &  .... &13.5$\pm$4.9\\
   & A.1      &115 &4.50 -- 4.07 &3.25 -- 1.55 &10.0 -- 24.0 &$-$0.19, 0.05   &$-$0.23, 0.02&16.8$\pm$4.8\\
   & A.2      & 22 &4.17 -- 4.00 &3.04 -- 1.50 &3.0 -- 10.0  &$-$0.29, 0.16   &{\bf $+$0.70, 0.00}&7.4$\pm$1.8\\
\hdashline
A  &all stars &210 &4.04 -- 3.84 &4.50 -- 1.00 &0.0 -- 11.0  &{\bf $-$0.57, 1.78$^{-19}$}&$+$0.06, 0.37& 3.7$\pm$2.0\\
   &V/IV      &132 &    ....     &4.50 -- 3.60 &0.0  -- 4.0   &$+$0.05, 0.59         &$-$0.32, 0.00&2.8$\pm$1.0
\\
   &III       & 6  &  ....      &3.59 -- 2.41 & .... & .... & .... & ....\\
   &II/I      &65  &  ....       &2.40 -- 1.00 &2.0 -- 11.0  &{\bf $-$0.76, 2.10$^{-13}$}  &$+$0.24, 0.05 &5.7$\pm$2.1
\\
\hdashline
F  &all stars &233 &3.89 -- 3.74 &5.00 -- 0.50 &1.0 -- 6.2   &{\bf $-$0.72, 2.83$^{-39}$}&{\bf $+$0.56, 1.85$^{-20}$} & 2.3$\pm$1.1\\
   &V/IV      &168 &  ....       &5.00 -- 3.60 &1.0 -- 3.8   &$-$0.37, 5.69$^{-07}$&{\bf $+$0.77, 9.04$^{-34}$}&1.8$\pm$0.5\\
   & III      &24  &             &3.59 -- 2.41 &             & ....                    &....& ....\\
   &II/I      &41  &  ....       &2.40 -- 0.50 &2.7 -- 6.2   &$-$0.23, 0.14 &$+$0.04, 0.82 &4.2$\pm$0.8\\
 \hdashline
G  &all stars &312 &3.80 -- 3.65 &5.00 -- 1.00 &0.2 -- 7.0   &{\bf $-$0.84, 0.00} &$+$0.04, 0.51   &1.4$\pm$1.0\\
   &V/IV      &203 &   ....      &5.00 -- 3.50 &0.2 -- 2.1   &{\bf $-$0.71, 1.85$^{-32}$} &{\bf $+$0.70, 9.27$^{-32}$} &1.0$\pm$0.4
\\
   &III       &69 &   ....      &3.49 -- 2.31 &1.0 -- 4.0   &$-$0.42, 0.00 &$+$0.35, 0.00 &1.6$\pm$0.6\\
   &II/I      &40  &   ....      &2.30 -- 1.00 &1.3 -- 7.0   &{\bf $-$0.46, 0.00} &{\bf $+$0.77, 6.20$^{-09}$}&3.2$\pm$1.4
\\%CP apart
   \hdashline
K  &V         &106 &3.71 -- 3.62 &5.00 -- 4.10 &0.2 -- 2.0   &$-$0.13, 0.17 &$-$0.11, 0.26 &0.6$\pm$0.3\\
\end{tabular}
\end{center}
\small
{\bf Notes.} \newline
$\rho$ and  $p$ -- the Spearman rank correlation coefficient and its significance. Significant,
moderate to strong correlations are highlighted in bold.\newline
"a'' --  for all but  B\&G spectral type, no statistical results for LC~III objects have been
provided due to low statistics;
"b'' -- to  simplify the analysis, small differences in the \Teff\ range appropriate for LC~V/IV and
LC~II/I objects has been neglected, and an average of the two is only provided.
\end{table*}

From the top panels of Fig.~\ref{fig3} it appears that photospheric micro-turbulence
of the sample O stars (symbols coloured in blue) tends to increase towards lower
gravity from generally subsonic (\vmean$=$8.0$\pm$4.1~\kms) to supersonic
(\vmean$=$14.3$\pm$2.8) velocities  with a borderline between the two regimes
observed at \logg$\approx$3.7~dex (i.e., at the value separating roughly LC~V/IV
from the LC~III/I stars, see \citealt{markova18}). Additionally, a tendency of
cooler dwarfs/subdwarfs to appear, on average, less affected by \vmic\  compared
to the hotter ones seems to emerge as well. While the former observation has
been statistically confirmed, no evidence of any significant correlation between
\vmic\ and \Teff\ has been derived (within each of the two LC subgroups or over
the complete sample; see corresponding data in Table~\ref{stat_prop}).

There are, at least, two points of potential concerns regarding the above outlined
results: first, since our O-star data sets exclusively originate from FW analyses,
it may well be that these findings are subjects of systematic uncertainties as
reported by  \citet{massey13}, and second, due to the general scarcity of more massive
O-stars close to the ZAMS  (see Sect.~\ref{distr_vmic}), our findings may not be
representative  for the complete O star regime.

To deal with the first issue,  we have contrasted the temperature distribution of
micro-turbulent broadening  of our O-stars  to analogous  results for O stars in the
Small Magelanic Clouds (SMC) derived by \citet{heap06} using the CMFGEN code (in the top panels of
Fig.~\ref{fig3}, symbols highlighted in pink). While the dependence of \vmic\ on
metallicity is still an open issue, the consistency between the two data sets
is reasonably good (no evidence of any systematic shifts) thereby allowing to
conclude that the \vmic\ properties of Galactic O-stars derived  by us are not
likely to be significantly affected by the particular modelling.

Regarding the second issue, from a direct comparison of the \logg and \Teff\ ranges 
covered by our O-star sample  to the ones theoretically predicted by  \citet{martins05a} (in
the top panels of Fig.~\ref{fig3} the red solid (DWs) and dashed (Gs/SGs) vertical
lines), it appears that we have only two objects with log\Teff$\gtrsim$4.6. This
result indicates that the \vmic properties of Galactic O stars derived by us are
representative only for objects cooler than  $\sim$40~kK only.

\paragraph{B-type stars}\label{spt_B}

The B-star  subsample consists of 323 objects  of  SpT from B0 to  B9 distributed
by LC as follows: $\sim$49\% are of LC~V/IV: $\sim$42\%  -- of LC~II/I, and 9\% 
-- of LC~III (see upper panel of Fig.~\ref{fig1})

From the second raw panels of Fig.~\ref{fig3}, we see that in spite of the
relatively large dispersion at a given \Teff\ and \logg\ (generally
$\lesssim$2$\sigma$ measurement error), the behaviour of \vmic\ is not chaotic
but follows a number of well defined patters. Specifically, at
3.70~dex$\le$\logg$<$4.50~dex, (LC~V/IV, solid circles; see \citealt{FM05, lefever10, nieva13})
typical velocity ranges from  zero up to $\sim$12~\kms with a tendency to decrease
toward a cooler temperature with values indistinguishable from zero at
\Teff$\le$4.2~dex.  At 3.30~dex$\le$\logg$<$3.70 (LC~III, black open circles) the 
photospheric micro-turbulence is, on average, larger than for the high gravity 
objects (\vmean=6.9$\pm$4.0~\kms vs.
\vmean=3.8$\pm$3.2~\kms) and appears to decline towards cooler \Teff\  as well.

Finally, at \logg$<$3.30~dex (LC~II/I;  red open circles; see \citealt{lefever07, markova08, MP08,
searle08}), two distinct \vmic\ subgroups clearly emerge with a borderline
between them seen at log\Teff$\approx$4.10-4.15~dex. On the hotter side, the
photospheric micro-turbulence is generally supersonic ($>$10~\kms) and tends
to increase with decreasing \Teff\ (Group~A.1); on the cooler one, it is,
on average, subsonic and seems to decrease with decreasing \Teff\ instead (Group~A.2).
(Warning: Due to low statistics, the results obtained for LC~III stars and
Gr.~A.2 objects should be considered with caution.)

\paragraph{AF-type stars}\label{spt_AF}

The number of A- and F-type stars in our analysis is 210 and 233, respectively,
with LC V/IV objects apparently dominating  over the LC~III/II/I ones: $\sim$63\%
vs. $\sim$37\%, respectively (upper  panel of Fig.~\ref{fig1} and the data
listed  in Column 3 of Table~\ref{stat_prop}).

From the third raw panels of Fig.~\ref{fig3} it appears that photospheric
micro-turbulence of the sample A stars  tends to strengthen toward lower gravities.
The relationship is found to be moderately strong and significant %at the 99.9\% confidence level
with LC~V/IV  objects (\logg$\ge$3.6~dex, solid circles; see e.g. \citealt{niemczura15}) 
indicating velocities, on average, a factor of two lower that those of LC~II/I ones
(\logg$\le$2.4~dex, open circles; e.g. \citealt{venn95, verdugo99}). Although less affected by
\vmic\ compared to A-type stars, the sample F-stars (fourth raw panels of
Fig.~\ref{fig3}) also provide evidence of a strong dependence of this quantity
on \logg\ with LC~II/I  objects (\logg$\le$2.4~dex, open circles; see Lyubimkov eta l. 2010, 2015)
having \vmean a twice lower than that of LC~V/IV ones (\logg$\ge$3.6~dex, solid circles;
see \citealt{niemczura15, takeda05b}).

Concerning the temperature behaviour of \vmic, for AF-type SGs this quantity is
temperature independent; for AF-type DWs, however,  a  weak  negative (SpT A) and
a very strong positive (F SpT)  correlation has been derived in perfect qualitative
agreement with previous findings from \citet{gray01,smalley04, takeda08, gebran14,
niemczura15}.

\paragraph{GK-type stars}\label{spt_GK}

The total number of G-stars in our analysis is  312 with 203 classified of
LC~V/IV  ($\sim$65\%), 69 -- of LC~III ($\sim$22\%), and 40 of LC~II/I
($\sim$13\%).
From the fifth raw panels of Fig.~\ref{fig3} we find that also in this
temperature regime low-gravity objects (LC~II/I, \logg$\le$2.3~dex, open circles;
see \citealt{lyubimkov10, lyubimkov15, takeda05a, takeda05b}) appear  more
strongly affected by \vmic\ compared to the high-gravity ones
(\logg$\ge$3.5~dex, LC~V/IV, solid circles): \vmean$=$3.2$\pm$1.4 vs. \vmean$=$1.0$\pm$0.4
\kms.  Additionally, within each LC subgroup this quantity is strongly
correlated with \Teff: the stronger  \vmic -- the cooler the stars
(Column 7\&8 of Table~\ref{stat_prop}).

Regarding the sample of K-type stars (a total of 106, all of LCV/IV with \logg$\ge$4.10~dex; see sixth-row panels of Fig.\ref{fig3}), although their \vmic\ values are generally low, they are on average significantly different from zero (\vmean=0.6$\pm$0.3). Moreover, \vmic\ appears to be independent of both \Teff\ and \logg.  Due to the narrow \logg/log\Teff\ range covered by our dataset, these findings should be interpreted with caution.
\begin{table*}
\footnotesize
\begin{center}
\caption[]{Statistical properties of photospheric micro-turbulence  of the
sample non-radial pulsators separated by the kind of oscillations, and of
the classical Cepheids and Red giants/supergiants divided into two mass/luminosity
subgroups.  }\label{stat_propP}
\tabcolsep1.0mm
\begin{tabular}{llrcccrllc}
\hline
SpT &type of  &\# of &log\Teff &\logg  &$\La$/$\La{_\odot}$&\vmic limits  &$\rho$, $p$    &$\rho$, $p$          &\vmean \\
    & puls    &stars &limits   &limits &limits             &[\kms]        &(\vmic, \logg) &(\vmic, log\Teff) & [\kms]\\
\hline
\hline\\
$\beta$ Cep &non-RPs &14&4.44 -- 4.30&4.15 -- 3.50&2.67 -- 3.45&0.0 -- 14.0&{\bf$-$0.78, 0.00}$^{a}$&$+$0.31, 0.28$^{a}$ & 6.1$\pm$4.0\\
reference stars&non-var MS B&112&4.43 -- 4.22&4.40 -- 3.20&2.67 -- 3.45&0.0 -- 14.0& $-$0.26, 0.00&$+$0.01, 0.89&5.6$\pm$3.6\\
\hline
SPBs genuine  &non-RPs &41  &4.34 -- 4.04  &4.30 -- 3.05 &1.59 -- 3.28 &2.0 -- 7.0 &$-$0.07, 0.66 & $-$0.44, 0.00 &3.9$\pm$1.7\\
SPBs suspected   &non-RPs&13  &4.13 -- 4.00  &3.60 -- 2.83 &1.79 -- 2.79 &2.0 -- 6.6 &$-$0.36, 0.23 &$+$0.31, 0.30 &3.7$\pm$1.3\\
SPBs total &non-RPs  &54& 4.34 -- 4.00&4.30 --2.83 &1.59 -- 3.28&2.0 -- 7.0&$-$0.14, 0.31&$-$0.40, 0.00&3.8$\pm$1.6\\
reference stars&non-var MS B&158&4.51 -- 4.00&4.57 -- 3.04&1.72 -- 3.27&0.0 -- 13.0&$-$0.16, 0.04&$+$0.33, 0.00&3.9$\pm$3.2\\
\hline
$\delta$ Scu &non-RPs&44 &3.97 -- 3.80&3.40 --4.32&0.42 -- 1.58 &1.2 -- 4.1 &$+$0.09, 0.57 &non-mnt &2.8$\pm$0.8\\
$\gamma$ Dor &non-RPs&52 &3.90 -- 3.78&4.50 -- 3.80&2.29 -- 2.29&1.3 -- 3.2 &$-$0.28, 0.04 &non-mnt&3.1$\pm$0.4\\
reference stars&non-var AF dws&320 &4.00 -- 3.85 &4.87 -- 3.85 &1.54 -- $-$0.4& 0.0 -- 4.0&$-$0.21, 0.23&non-mnt&\\
\hline
Cepheids, total & RPs &199 &3.83 -- 3.69&2.60 -- 0.50&2.00 -- 3.75&2.4 -- 6.5 &$+$0.03, 0.68 &$+$0.04, 0.56 &4.1$\pm$0.8\\
Ceph \Minit$<$9Msun& -- &164&3.83 -- 3.69&2.60 -- 1.18&2.00 -- 3.08&2.4 -- 6.3 &$+$0.09, 0.24&$+$0.07, 0.34 &4.0$\pm$0.8\\
Ceph \Minit$>$9Msun& -- &35 &3.82 -- 3.69&1.55 -- 0.50&3.11 -- 3.75&3.0 -- 6.5 &$-$0.03, 0.88&$-$0.08, 0.66 &4.2$\pm$1.0\\
RGs          &var &215 &3.74 -- 3.60&3.75 -- 1.00&2.81 -- 0.44&0.7 -- 2.1&{\bf $-$0.42, 2.58$^{-10}$} &$-$0.24, 0.00&1.5$\pm$0.2\\
RSGs        &var &7 &3.60 -- 3.59  &1.00 -- 0.30 &3.45 -- 2.81   &1.1 -- 1.6 & .... & .... &1.3$\pm$0.2\\
\end{tabular}
\end{center}
\small
{\bf Notes.} \newline
$\rho$ and $p$ -- the Spearman rang correlation coefficient and its significance.
Significant, moderate to strong correlations are highlighted in bold.\newline
reference stars = sample non-variable stars of similar \Teff, \logg\ and  spectral luminosity;
SPBs = Slowly Pulsating B stars; SPBs suspected = stars which lie outside the
instability strip for $g$ mode oscillations but posses photometric and \vmic\
properties similar to those of the genuine SPBs; RPs = radial pulsators;
non-RPs~=~non-radial pulsators; var~=~variable stars; non-mnt~=~non-monotonic
behaviour;
"a" -- due to lower statistics these results should be considered with caution.
\end{table*}

\subsection{Pulsating stars and objects in the Red giant phase}\label{vmic_puls}

There is both direct and indirect evidence that stellar pulsations arising from coherent gravity 
($g$) or pressure ($p$) mode oscillations can significantly affect the macro-turbulent velocity (\vmac) derived for main-sequence B-stars using the Fourier Transform (FT) or FT combined with Goodness-of-Fit (FT+GOF) methods (\citealt{aerts14, SS17}, and references therein). In some cases, these modifications can reach amplitudes of several tens of \kms.
\begin{figure*}
\begin{center}
{\includegraphics[width=7.18cm,height=3.64cm]{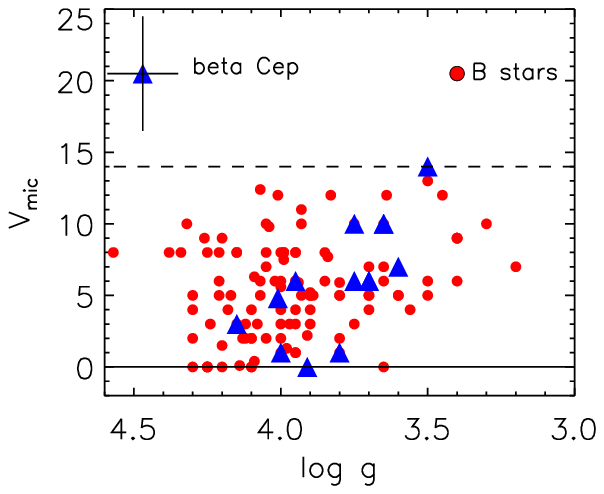}}
{\includegraphics[width=7.18cm,height=3.64cm]{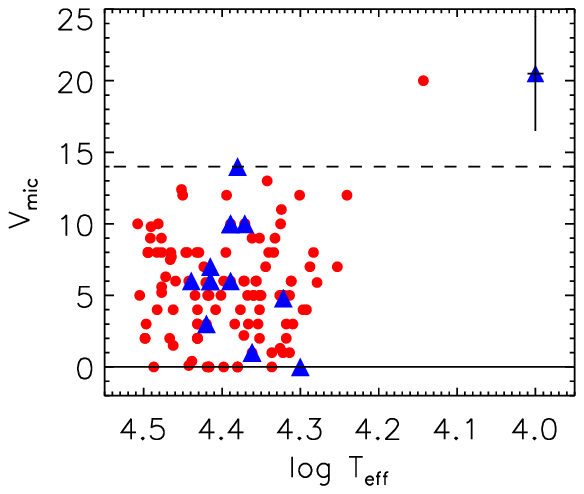}}\\
{\includegraphics[width=7.18cm,height=3.64cm]{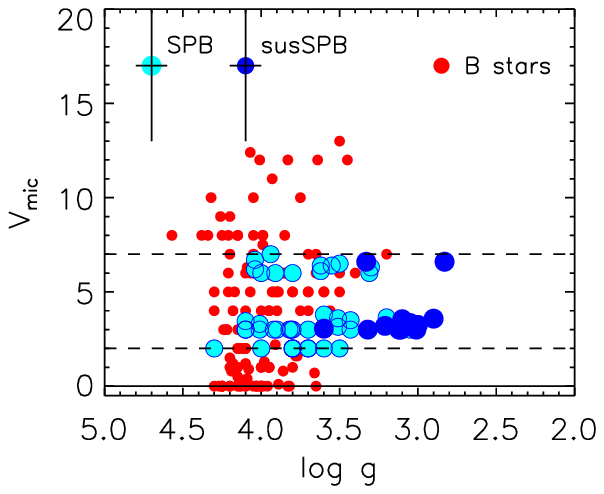}}
{\includegraphics[width=7.18cm,height=3.64cm]{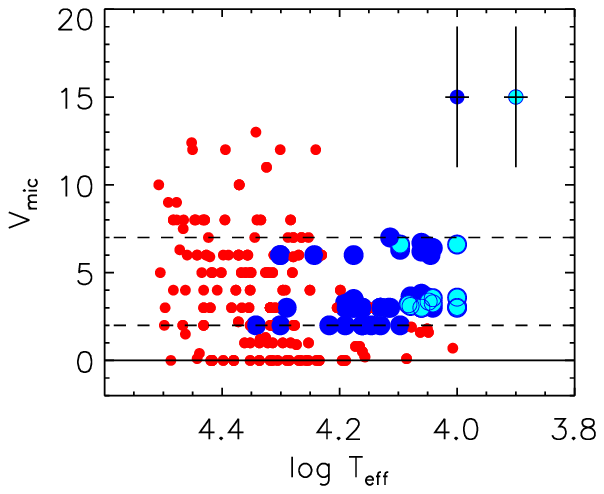}}\\
{\includegraphics[width=7.18cm,height=3.64cm]{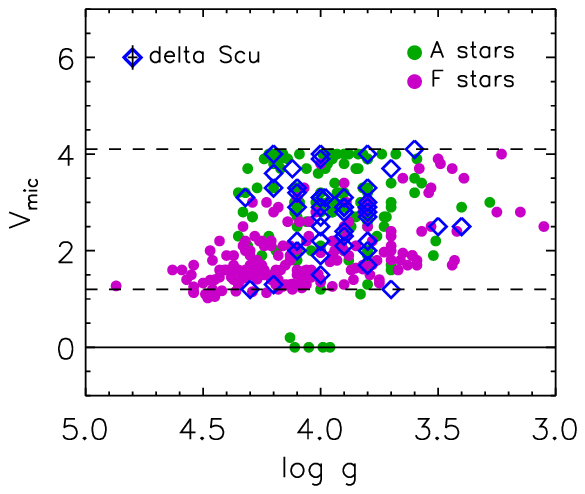}}
{\includegraphics[width=7.18cm,height=3.64cm]{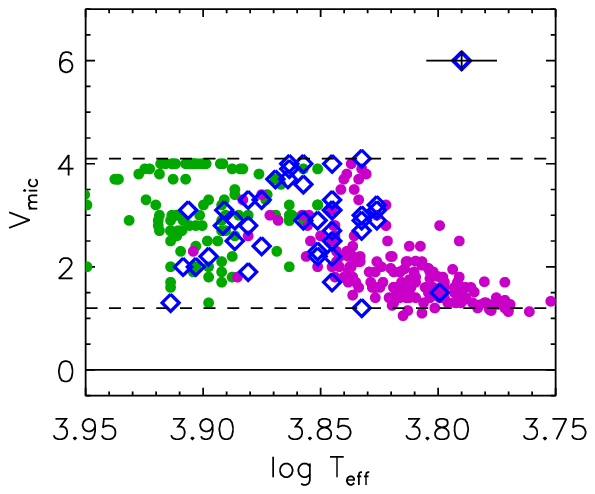}}\\
{\includegraphics[width=7.18cm,height=3.64cm]{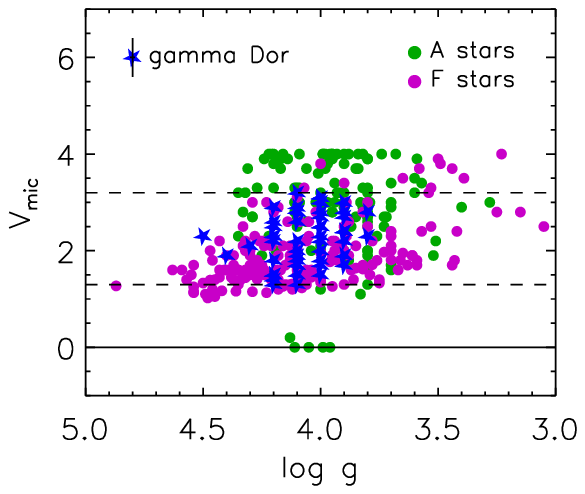}}
{\includegraphics[width=7.18cm,height=3.64cm]{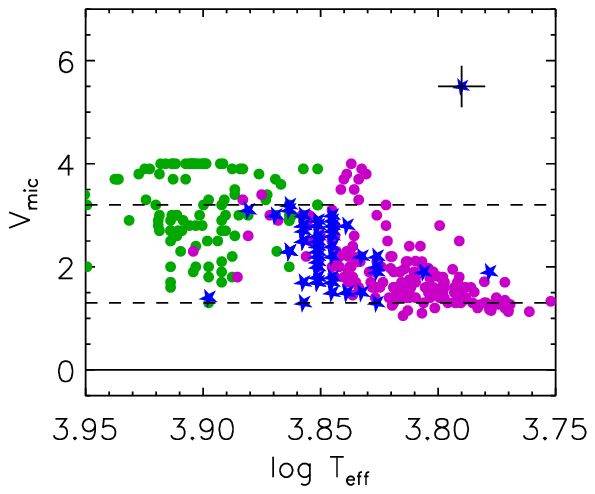}}\\
{\includegraphics[width=7.18cm,height=3.64cm]{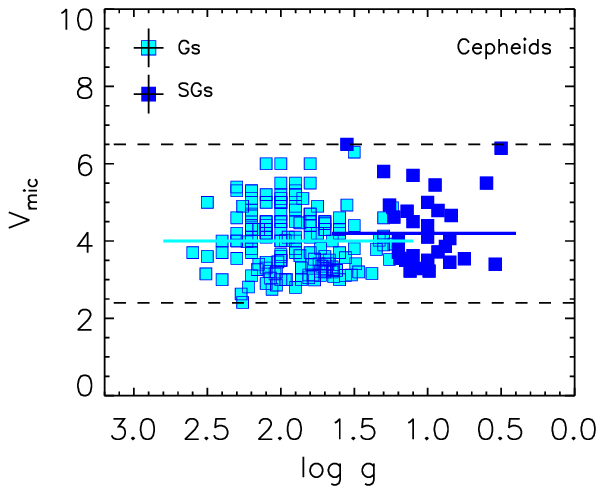}}
{\includegraphics[width=7.18cm,height=3.64cm]{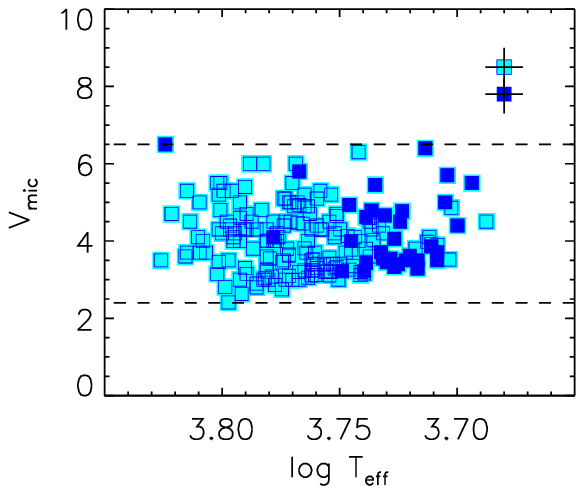}}\\
{\includegraphics[width=7.18cm,height=3.64cm]{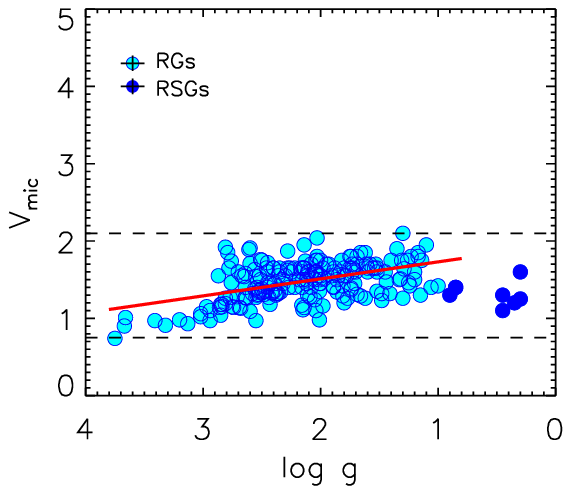}}
{\includegraphics[width=7.18cm,height=3.64cm]{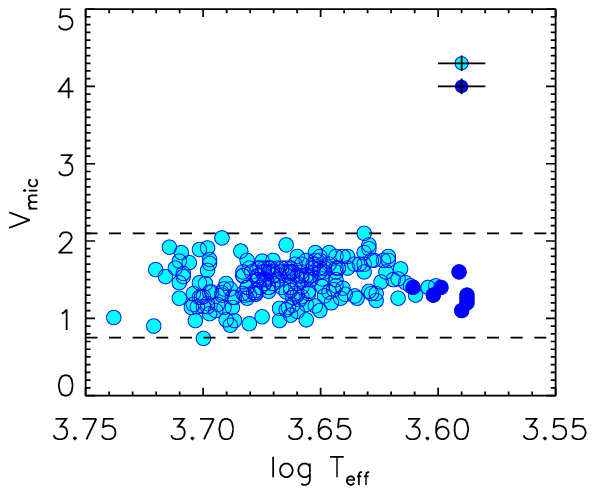}}
\caption{ Photospheric micro-turbulence of the sample non-radial pulsators
(separated  by the type of oscillations), and of the classical Cepheids and Red
Giants/Supergiants  (symbols highlighted in light/dark blue) as a function of 
\logg (left) and \Teff (right). For  the case of non-RPs analogous results for 
non-pulsating stars of similar \Teff, \logg\ and $\La$/$\La{_\odot}$ (colour coded 
as indicated in Fig.~\ref{fig3}),  are also provided for a direct comparison. In 
each panel the horizontal dashed lines indicate the upper/lower limits to \vmic\. In
the left fifth raw panel,  the light and dark blue lines represent the mean 
\vmic, averaged within the  subgroups of low- and high-mass Cepheids, respectively; 
in the bottom left panel, the red line indicates the least square 
fit to \vmic\ of the sample RGs  thereby highlighting the deviation of RSGs from 
this trend. For more information see
Appendix~\ref{A}.}
\label{fig4}
\end{center}
\end{figure*}
Since both line-profile shapes and line equivalent widths (EWs) are predicted to vary throughout a pulsation cycle (see, e.g., Fig.7 in \citealt{aerts14}), and our analysis is based entirely on data obtained via line profile fitting (LPF) and EW measurements (Sect.\ref{vmic_data}), it is plausible that such oscillations may influence the observed micro-turbulent velocities.

To explore this possibility, we compiled a database of literature values for \Teff, \logg, and \vmic\ for approximately 160 Galactic stars identified as $p$- and $g$-mode pulsators of various types ($\beta$ Cephei, Slowly Pulsating B-stars, $\gamma$~Doradus, and $\delta$~Scuti variables).
\footnote{Although more massive and evolved stars are also known to exhibit pulsational instability (see, e.g., \citealt{godart17}, \citealt{haucke18}, and references therein), we focus here on intermediate- and low-mass non-radial pulsators on the main sequence to simplify the analysis.}
To this sample, we added 199 classical Cepheids (radial pulsators) and 222 red giants/supergiants (RGs/SGs), representing late evolutionary stages on the red-ward path of the Hertzsprung-Russell diagram, for completeness.

The primary results of the \vmic\ analysis for these variable stars are shown in Fig.\ref{fig4} and summarized in Table \ref{stat_propP}. Further discussion is provided in Appendix~\ref{A}, where Fig.~\ref{figA1} shows the location of these stars in the sHR diagram. Key findings are highlighted below. Note that for non-radial pulsators (non-RPs), we conducted a comparative analysis against non-pulsating reference stars with similar \Teff, \logg, and $\La/\La_{\odot}$ to assess any differences in \vmic\ behaviour. Also note that due to certain limitations — such as heterogeneity in pulsation types, reliance on snapshot rather than time-series data, limited sample sizes, and incomplete coverage in \Teff\ and \logg — these results should be interpreted as indicative rather than conclusive.

\textbullet\ All types of variable stars in our database show photospheric micro-turbulence values that are, on average,  significant, i.e., above their respective uncertainties (Column 10 of Table~\ref{stat_propP}).

\textbullet\ The \vmic\ properties of non-RPs are generally similar to those of the corresponding reference stars, with SPBs as a possible exception ( see first to fourth rows of Fig.\ref{fig4} and Columns 7–10 in Table\ref{stat_propP}). These findings suggest that pulsations are unlikely to play a dominant role in setting \vmic\ in $\beta$ Cephei, $\gamma$~Doradus, and $\delta$~Scuti stars. However, further investigation is required to determine whether high-order $g$-mode oscillations in SPBs significantly affect their photospheric EW profiles.

\textbullet\ For classical Cepheids, \vmic\ tends to increase with decreasing \Teff\ and \logg\ (and thus increasing \Minit). While this trend is qualitatively consistent with that observed for non-variable stars (Sect.\ref{vmic_spt}), it is not statistically confirmed (see Table\ref{stat_propP} and the two horizontal lines in the left fifth-row panel of Fig.~\ref{fig4}). This may be due to selection effects and large intrinsic dispersion in \vmic\ at given \Teff\ and \logg, likely arising from the combined use of snapshot and phase-averaged data \citep{proxauf18, luck18b}. A more robust analysis using exclusively phase-averaged data across the entire instability strip is necessary to conclusively determine whether higher-mass Cepheids are more affected by micro-turbulence.

\textbullet\ For red giants, \vmic\ increases with decreasing surface gravity, while red supergiants (RSGs) appear less affected than expected based on their \Teff\ and \logg\ (see red line in the bottom left panel of Fig.~\ref{fig4}). These results could indicate that RSGs exhibit different \vmic behaviour compared to RGs - a plausible outcome given the significantly different surface conditions of these two stellar groups \citep[see e.g.][]{GBP}. However, due to the small number of RSGs in our sample, this conclusion remains tentative.

Based on the analyses in this and the previous subsections, two main conclusions emerge: First, for all stars in our sample, regardless of spectroscopic or pulsational characteristics, photospheric micro-turbulence is significant — i.e., above measurement uncertainty. Second, within each SpT, \vmic\ depends meaningfully on \Teff\ and \logg, but the relationship lacks a uniform pattern that would allow for a deeper understanding of its origin when studied within a single SpT. These findings caution against neglecting \vmic\ or assuming a fixed, ad hoc value across all stars of a given SpT. Such simplifications may introduce systematic errors into derived chemical abundances, particularly for O-type stars. For example, assuming an ad hoc \vmic=10~\kms\ can lead to systematic overestimation of log~(N) by $\sim$0.04–0.08~dex (CMFGEN) and $\sim$0.07–0.14~dex (FASTWIND) in supergiants, and an underestimation of similar magnitude in dwarfs — especially toward the cooler end of the temperature regime.

\section{Micro-turbulence across the sHR diagram. }\label{vmic_shrd}
\subsection{General features}\label{gen_features}

Although diverse, the behaviour of \vmic\ across the sample stars exhibits one consistent trend: within each spectral type (SpT), low-gravity stars (LCII/I) show systematically higher micro-turbulent broadening than high-gravity stars (LCV/IV). This pattern, previously hypothesized by \citet{gray01}, \citet{bouret05}, \citet{lyubimkov10}, \citet{holgado18}, and \citet{markova18} — though based on smaller samples within specific SpTs — suggests that, on a global scale, photospheric micro-turbulence may depend on stellar mass and evolutionary stage.

With this in mind, we constructed a spectroscopic HR diagram for all sample stars (both pulsating and non-pulsating), where each star is represented by a circle colour-coded and size-scaled according to its corresponding \vmic\ (see legend in Fig.\ref{fig2}).
Although somewhat patchy (see Sect.\ref{distr_vmic}), this \vmic map benefits from the large and diverse sample, allowing for the emergence of several notable trends. In particular, the map suggests that photospheric micro-turbulence tends to decrease along a trajectory from the top-left to the bottom-right of the diagram — that is, with decreasing $\La/\La_{\odot}$ (=\Teff$^4$/\textit{g}) and \Teff. This trend is supported by the data presented in Fig.\ref{fig5}, which highlight three key findings:

1. Despite significant scatter at fixed $\La/\La_{\odot}$ (see below), \vmic\ generally decreases with decreasing spectroscopic luminosity and hence lower initial stellar mass (\Minit).

2. At similar $\La/\La_{\odot}$ values, stars of different SpT exhibit significantly different \vmic\ values — typically differing by more than 2--3$\sigma$ (see Column 7 of Tables~\ref{spdat} and \ref{photdat}). In general, later-type stars show lower \vmic\ than earlier-type ones. Since later SpTs at a given $\La/\La_{\odot}$ correspond to more evolved objects, this finding suggests that micro-turbulence may be influenced not only by \Minit but also by the degree of stellar evolution.

3. For stars with log($\La/\La_{\odot}$) $\gtrsim$ 3.1 dex, a clear upper detectability limit to \vmic\ becomes apparent.
\begin{figure}
\resizebox {\hsize}{!}{\includegraphics{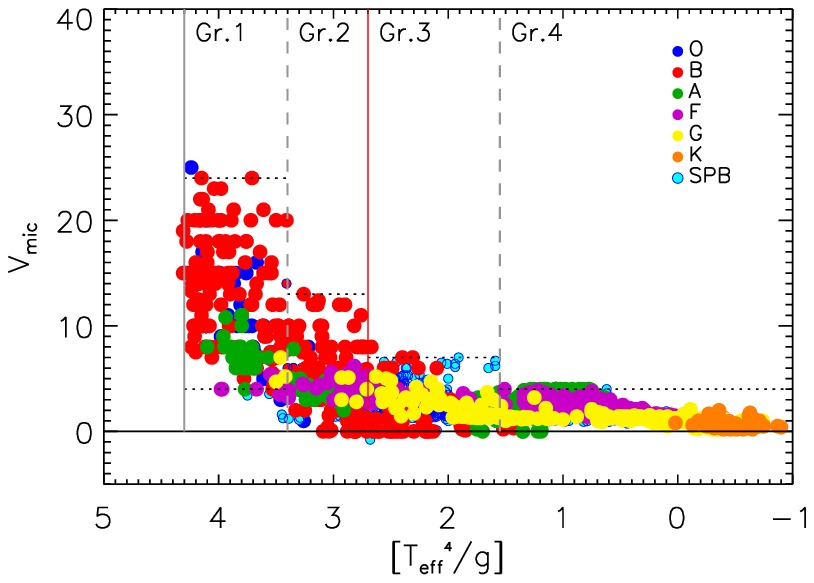}}
\caption{Photospheric micro-turbulence of the sample stars  (pulsating and
non-pulsating) as a function of log$\La$/$\La{_\odot}$ (=[\Teff$^{4}$/$g$]).
The data points are color coded according to their SpT (as indicated in the legend) with  Slowly
Pulsating B-stars (SPBs) highlighted in light blue. Vertical lines divide the sample into four spectral luminosity subgroups, depending on
the maximum \vmic\ velocity achieved  by  the corresponding stars (horizontal
dotted lines), with the red one indicating the approximate limit between the high
and low mass regimes (for more information see text).
}
\label{fig5}
\end{figure}

\subsection{Dependence on stellar initial mass and evolution}\label{vmic_mass}
\begin{figure*}[t]
{\includegraphics[width=8.8cm,height=6.2cm]{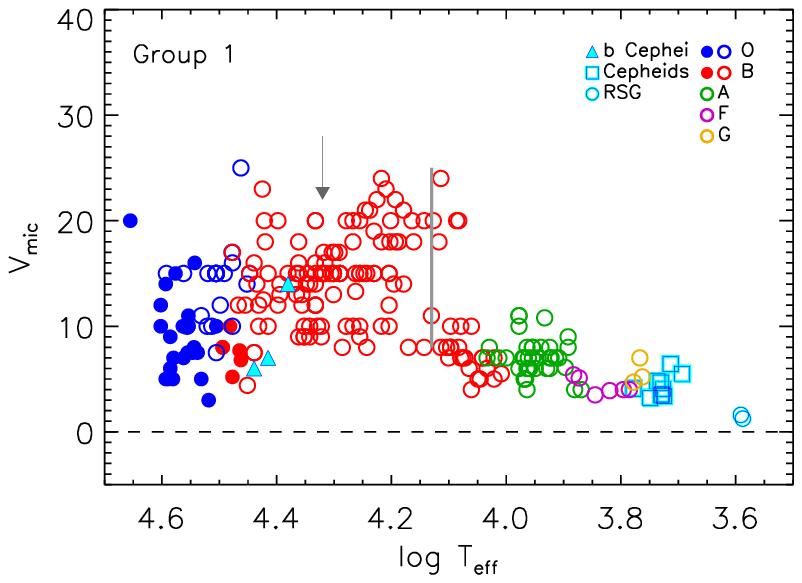}}
{\includegraphics[width=8.8cm,height=6.2cm]{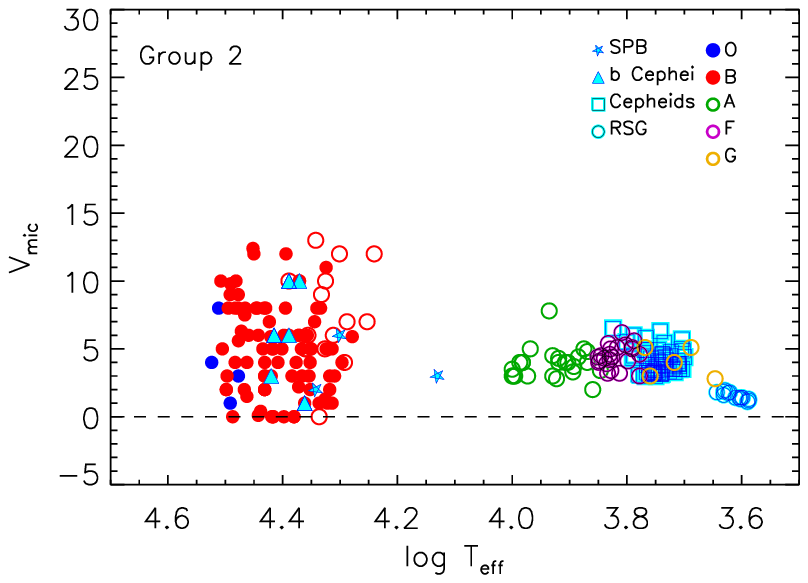}}\\
{\includegraphics[width=8.8cm,height=6.2cm]{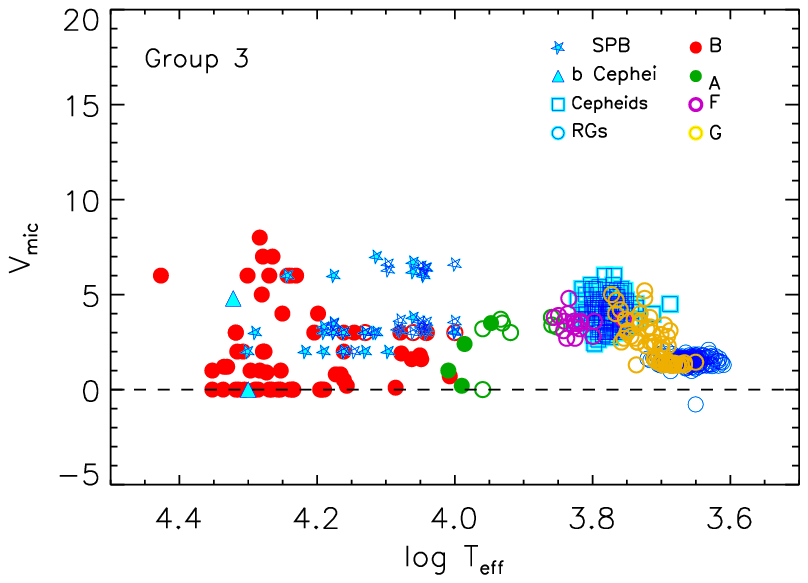}}
{\includegraphics[width=8.8cm,height=6.2cm]{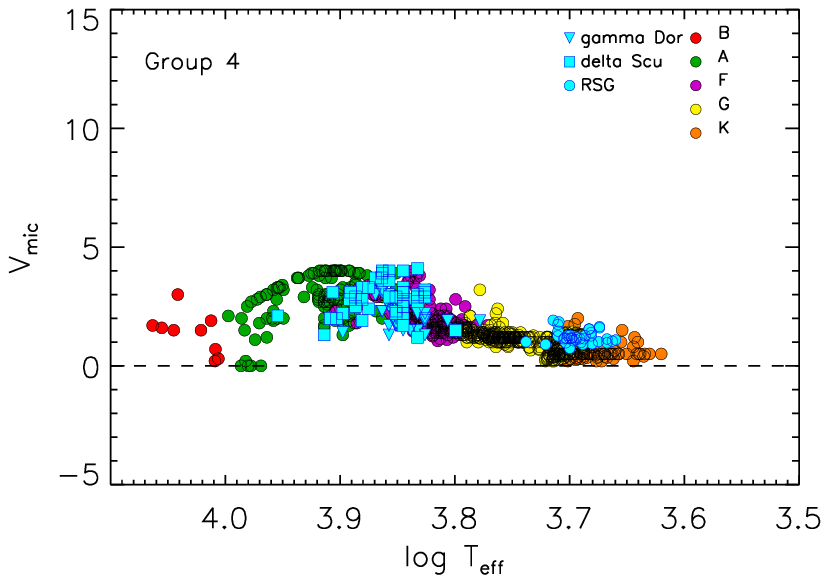}}
\caption{The \vmic -- \Teff\ distribution of the sample stars divided into
four spectral luminosity subgroups as defined in Sect.\ref{vmic_mass}.
Symbols and colours as indicated in the legend with LC~V/IV and LC~III/I
targets highlighted by solid and open circles, respectively. In the top left
panel the steep drop in \vmic is highlighted by a gray vertical line.
For high-mass stars (\Minit$\ge$8\Msun, Gr.~1\&2) the uncertainty
in the determination of \vmic\ is typically  between 1 and 5\kms;
for low-mass ones (\Minit$\le$8\Msun, Gr.~3\&4) it is generally below 1~\kms 
 (for more details see text).
}
\label{fig6}
\end{figure*}

To investigate the potential dependence of photospheric micro-turbulence on initial mass (\Minit) and stellar evolution, we adopted a two-step approach. First, to reduce the confounding influence of $\La/\La_{\odot}$, we divided the sample into four spectral luminosity subgroups based on the maximum observed \vmic  values  achieved by the corresponding stars (see horizontal dotted lines in Fig.\ref{fig5}). Second, within each luminosity subgroup, we examined the behaviour of \vmic\ as a function of \Teff, clearly distinguishing between stars of different SpT, LC, and pulsation types. The main results of this analysis are illustrated in Fig.\ref{fig6}, with detailed commentary provided below.

\textit{Group 1 (\vmic$\lesssim$24~\kms, 3.4$\lesssim$log($\La/\La_{\odot}$)$\lesssim$4.3; roughly corresponding to \Minit$\gtrsim$15\Msun)}
As seen in the upper left panel of Fig.~\ref{fig6}, Group 1 is primarily composed of more massive OB-type stars and their evolutionary descendants — luminous AFG-type supergiants. A small number of very luminous classical Cepheids and red supergiants (RSGs) are also present in this group.

While the photospheric micro-turbulence in Group 1 stars appears to depend on temperature, the relationship 
is non-monotonic. It begins with a gradual increase from subsonic (\vmic$\approx$5~\kms) to highly 
supersonic velocities  (\vmic$\approx$25~\kms) over the range 4.6$\lesssim$log\Teff$\lesssim$4.15, followed by 
a sharp decline to \vmic$\sim$10~\kms at log~\Teff$\approx$4.15 (marked by a solid vertical line), and a more 
moderate decrease thereafter. Late-B and A-type supergiants are more strongly affected than their more evolved 
FG-type counterparts (including Cepheids), with average values of 7.4$\pm$1.8~\kms and 7.0$\pm$1.8~\kms versus 
4.3$\pm$0.7~\kms and 4.3$\pm$1.0~\kms, respectively. There is suggestive evidence of a further decrease in \vmic\ 
toward RSGs, though the sample size is too small for firm conclusions.

A noteworthy detail is the potential local minimum in \vmic\ (indicated by a vertical arrow), which is intriguingly close to the location of the bi-stability jump identified by \citet{MP08} at log~\Teff$\approx$4.3~dex. This proximity raises the possibility that micro-turbulence in this mass and temperature regime could be influenced, either directly or indirectly, by changes in stellar wind properties.

The sudden drop in \vmic\ at log~\Teff$\approx$4.15 may be explained by its proximity to the transition between the horizontal and temperature-dependent segments of the Humphreys-Davidson (HD) limit on the sHR diagram (Fig.\ref{fig2}, upper panel, dark gray lines). As the HD limit is typically associated with intense mass loss that halts the evolution of stars with \Minit$\gtrsim$25~\Msun\ toward cooler temperatures, this context may account for the observed drop. While this hypothesis is supported by the fact that most B-type SGs with \vmic$\le$10~\kms have \Minit$<$20--30\Msun, the presence of a few outliers with much higher \vmic\ suggests that other contributing factors may also be at play.

\textit{Group 2 (\vmic$\le$14\kms, 2.7$<$log($\La/\La_{\odot}$)$\le$3.4; \Minit$\sim$8--15\Msun)}
This subgroup includes a small number of O-type dwarfs, many B-type main-sequence (MS) stars (including nearly all $\beta$~Cephei variables and two SPBs), a large number of massive Cepheids and AFG-type SGs, and a limited number of RSGs. Due to the Hertzsprung gap, we lack post-MS B stars within 4.25$\lesssim$log\Teff$\lesssim$4.0 (see Fig.\ref{fig2}).

The upper-right panel of Fig.\ref{fig6} shows that \vmic\ tends to decrease with decreasing \Teff\ (and thus more advanced evolutionary stages). However, this trend is not statistically confirmed. MS B-stars have \vmic\ values comparable to those of evolved AFG-type SGs (including massive Cepheids): average values are 5.7$\pm$3.3~\kms vs. 3.9$\pm$1.1~\kms, 4.5$\pm$0.8~\kms, and 4.2$\pm$1.0~\kms, respectively. In contrast, RSGs in this group exhibit significantly lower \vmic: \vmean=1.52$\pm$0.27~\kms, with a potential drop near log~\Teff=3.65--3.70~dex.

\textit{Group 3 (\vmic$\le$7~\kms, 1.6~$<$log($\La/\La_{\odot}$)$\le$2.7; \Minit$\sim$3--8\Msun)}
Group 3 includes many low-mass B-type dwarfs and SPBs, a limited number of MS A-type stars, numerous post-MS AFG-type objects (including Cepheids), and a large sample of red giants (169 stars).
\footnote{Stars with \Minit$\sim$3--8\Msun may experience blue loop evolution, so those with similar \Teff\ and \logg\ may be at different evolutionary stages.}

Although lower in amplitude than in Group 2, \vmic\ in Group 3 behaves similarly — relatively high and approximately temperature-independent above log~\Teff$\gtrsim$3.7~dex. The average values are: 2.11$\pm$2.21~\kms (B-type dwarfs), 3.83$\pm$1.60~\kms (SPBs), 3.25$\pm$0.46~\kms (A-type), 3.49$\pm$0.52~\kms (F-type), and 4.0$\pm$0.8~\kms (low-mass Cepheids). RGs in this group show the lowest \vmic\ values: \vmean~=1.50$\pm$0.27~\kms. Interestingly, the transition from high to low \vmic\ appears again around log~\Teff=3.65--3.70~dex, as seen in Group~2.

\textit{Group 4 (\vmic$\le$4\kms, 1.0~$<$log($\La/\La_{\odot}$)$\le$1.6; \Minit$\sim$1--3\Msun)}
This group is populated primarily by MS AFGK-type stars (including $\gamma$~Doradus and 
$\delta$~Scuti variables), and a smaller number of post-MS FG-type stars and RGs (see Figs.\ref{fig2} and \ref{figA1}).

Though \vmic\ is small (below 4~\kms), it remains significant and displays a non-monotonic temperature dependence. A clear peak is seen at 3.85~$\lesssim$log\Teff$\lesssim$3.95, with a gradual decline toward both higher and lower temperatures. Near log\Teff~=4.0 and 3.7dex, \vmic\ approaches zero (see lower-right panel of Fig.\ref{fig6}). This pattern is consistent with previous findings for AF-type dwarfs (Sect.\ref{non_puls}), and our results extend the observed decline into the cooler GK-dwarf and RG regimes.

To summarize, on a global scale photospheric micro-turbulence depends non-monotonically on $\La/\La_{\odot}$ (a proxy for initial mass) and \Teff\ (a proxy for evolutionary phase among stars of similar \Minit). The analysis reveals at least two broad maxima in \vmic: a stronger one among cooler O- and B-type SGs (\Minit$\gtrsim$15\Msun, 4.15$\lesssim$log\Teff$\lesssim$4.45~dex), and a weaker one among AF-type dwarfs (Group 4, 3.8$\lesssim$log\Teff$\lesssim$4.0~dex). These results suggest that \vmic\ is linked to one or more physical processes whose activity depends on mass and temperature, likely in a manner that reflects the trends observed in the empirical \vmic map.

\section{Discussion}\label{discussion}

In this section, we present several examples demonstrating how our observational database can serve as a complementary tool to evaluate theoretical scenarios proposed to explain specific phenomena observed in hot massive stars in the Milky Way. These include, for example, the hypothesized connection between micro-turbulence and stellar convection (Sect.\ref{vmic_conv}); the observed deficit of slow rotators and stars with very low macro-turbulent broadening (Sects.\ref{vmic_vrot} and \ref{vmic_vmac}, respectively); and the longstanding discrepancy between spectroscopic and evolutionary mass estimates for O-type stars (Sect.~\ref{mass_discrepancy}).
\begin{figure}[ht]
\resizebox{\hsize}{!}{\includegraphics{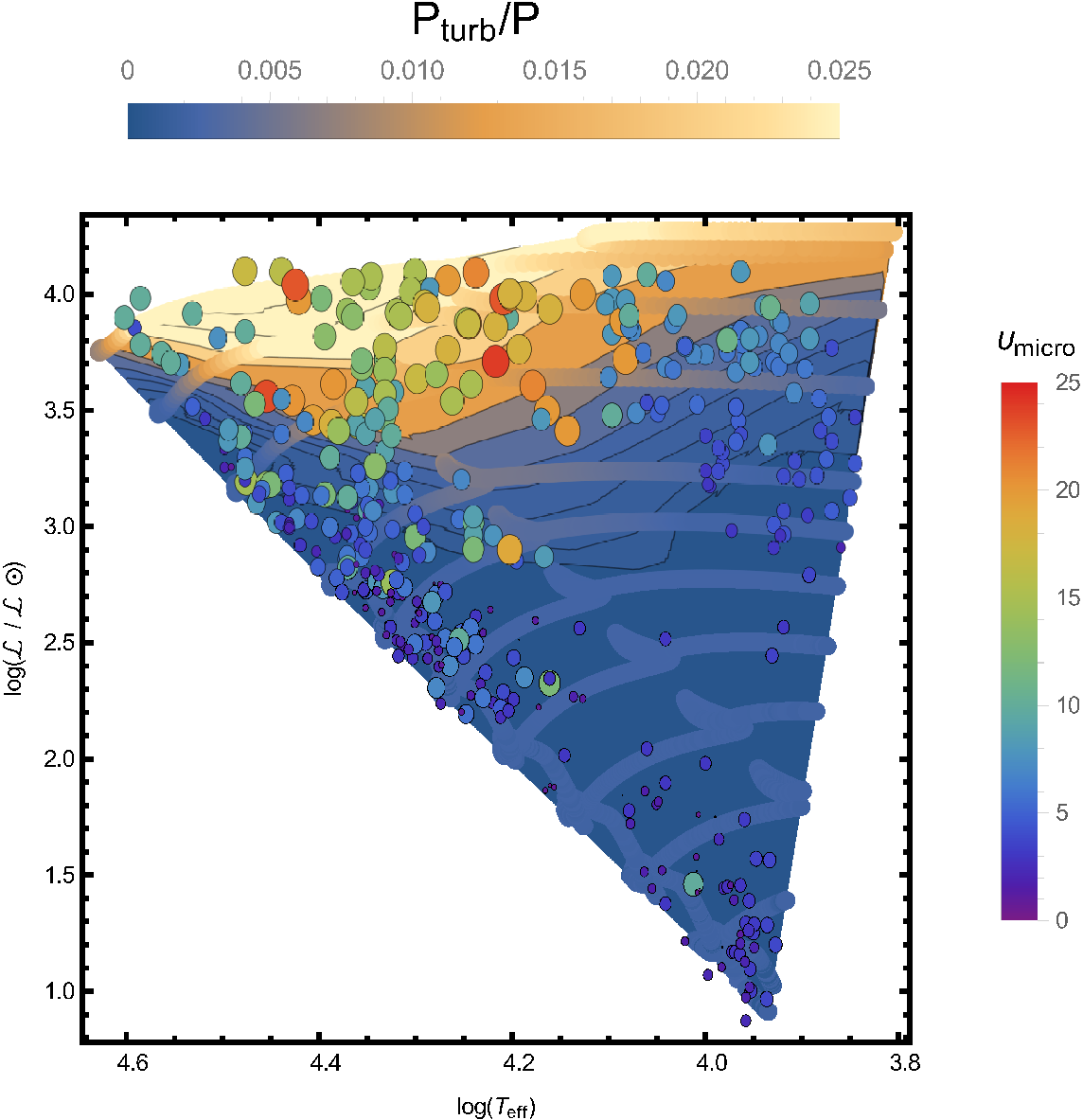}}\\
\resizebox{\hsize}{!}{\includegraphics{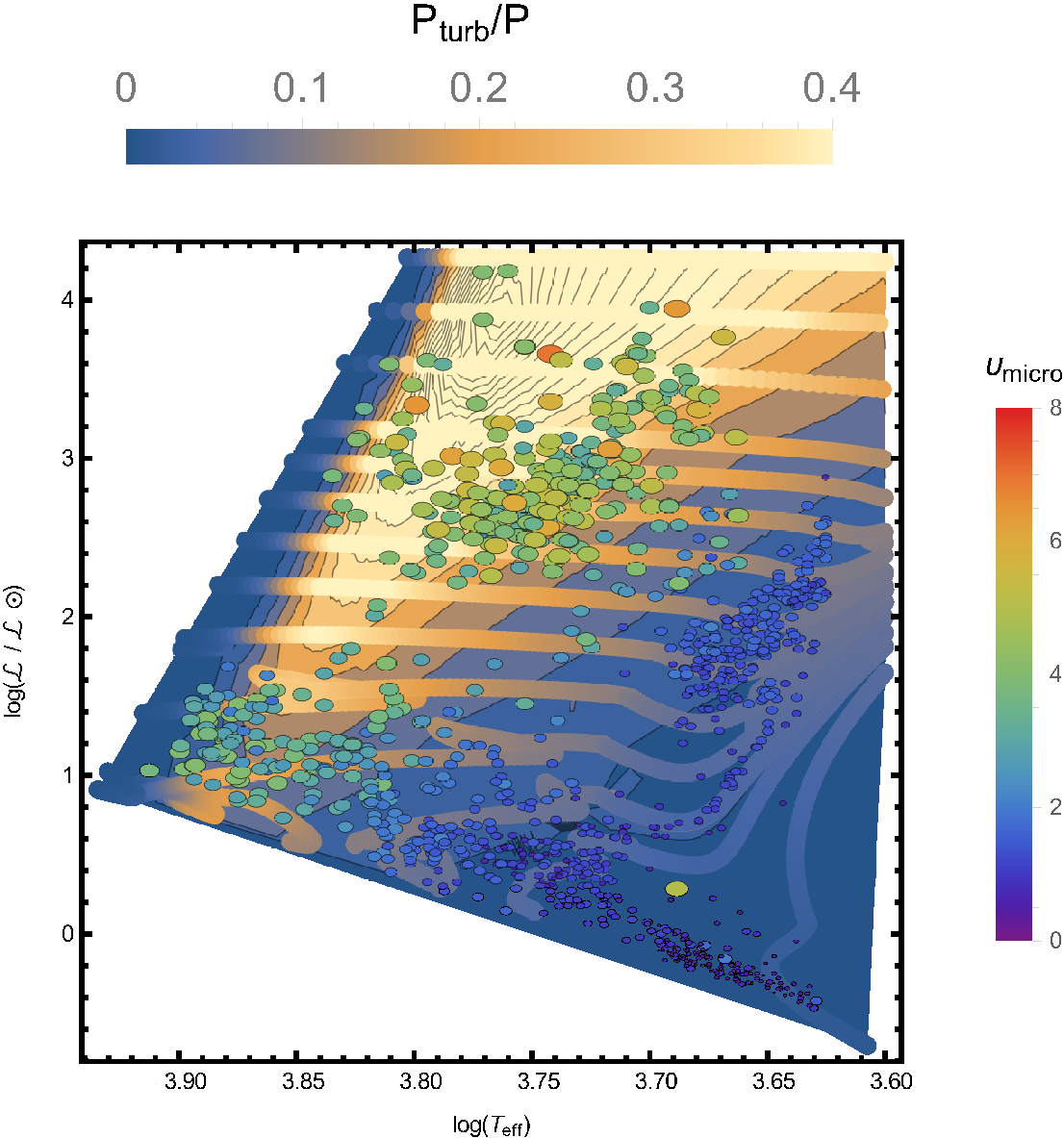}}
\caption{Spectroscopic HR diagram with coloured region representing the maximum ratio of turbulent pressure to total pressure in stellar envelopes as derived from 1D stellar evolution models. Over-plotted are photospheric micro-turbulent
velocities of the sample stars colour-coded as indicated in the legends (preliminary results).}
\label{fig7}
\end{figure}

\subsection{Micro-turbulence and stellar convection}\label{vmic_conv}

A physical connection between photospheric micro-turbulence and stellar convection in
low-mass stars was first proposed by \citet{edmunds78}, and later on confirmed
theoretically for the Sun by \citet{asp20a,asp20b}. For hot, massive stars,
\citet{cantiello09} similarly noted a correlation between observed micro-turbulent
velocities in OB stars at varying metallicity and the predicted average convective
velocities in the iron convection zone (FeCZ).

More recently, \citet{grassitelli15a, grassitelli15b} reported a close connection
between the empirical macro-turbulent velocity (\vmac) of a large sample of Galactic
stars and the maximum turbulent pressure fraction (MTPF) in non-rotating solar-metallicity
models. This result was confirmed by \citet{cantiello21}, who further showed that
the properties of subsurface convection correlate with both the timescale and the amplitude
of stochastic low-frequency photometric variability.
\begin{table*}[t]
\caption{Micro-turbulence and the deficit of slow rotators for Galactic stars
of various parameters separated by  spectral type (OBAFGK), luminosity class
(dw -- dwarfs; g -- giants; sg --supergiants) and initial stellar mass: Group~1
-- \Minit$\gtrsim$15~\Msun; Group~2 -- 8\Msun$\lesssim$\Minit$\lesssim$15\Msun;
Group~3 -- 3\Msun$\lesssim$\Minit$\lesssim$8~\Msun; Group~4 --
3\Msun$\lesssim$\Minit$\lesssim$1\Msun.
 }\label{vsini_vmic}
\footnotesize
\tabcolsep0.7mm
\begin{tabular}{l|ccccccccc|ccc|ccc|cc}
\hline
   \multicolumn{1}{l}{parameter} &
  \multicolumn{9}{c}{Group 1 } &
  \multicolumn{3}{c}{Group 2} &
  \multicolumn{3}{c}{Group 3} &
  \multicolumn{2}{c}{Group 4} \\
  \hline
  \multicolumn{1}{c}{} &
  \multicolumn{2}{c}{O dw} &
  \multicolumn{2}{c}{O g/sg} &
  \multicolumn{3}{c}{B g/sg} &
  \multicolumn{1}{c}{A }&
  \multicolumn{1}{c}{F } &
  \multicolumn{2}{c}{ B} &
  \multicolumn{1}{c}{AFG }&
  \multicolumn{2}{c}{B}&
 \multicolumn{1}{c}{FG} &
 \multicolumn{1}{c}{AF}&
 \multicolumn{1}{c}{GK}
 \\
 in \kms &hot & cool & hot & cool & hot & inter & cool &sg &sg &dw &g/sg & g/sg &dw &g &g &dw&dw \\
  \hline
  \hline
  \vmean     & ... & 8$^{a}$ &... &18$^{a}$&18$^{a}$& 15/20$^{a}$ &7.4(1.8) & 7.0(1.8)&4.3(1.1)&3.8(3.3) & ... &4.5/3.9/3.2 &2.1(2.2) &3.8(1.6)$^{b}$ &3.2(0.9) &3.0(2.4) &0.8(0.1)\\
  \vsinmean &60 &10   &65 &30  & 30   & 25/30  &15   & $\lesssim$5  &$\lesssim$5&$\lesssim$5   &20  &$\lesssim$5 &$\lesssim$5 &12 &$\lesssim$ 5 &9 &$\lesssim$2\\
 \hline
\end{tabular}
\\
\small
{\bf Notes.}
\vmean  -- mean value of \vmic\  averaged over the corresponding mass and temperature
regime and its standard error (in brackets); \vsinmean -- a ''by eye"  average over
nearby stars of similar upper detectability limit to \vsini; ''a" -- a "by eye" estimate;
''b" - an estimate corresponding to the sample SPBs
\end{table*}

Encouraged by the hypothesis that subsurface convection may drive small-scale velocity
fields in the stellar photosphere, we compared the \vmic\ map shown in Fig.~\ref{fig2}
with the predicted distribution of the MTPF from
\citet{grassitelli15a, grassitelli15b} (see their Fig.~4). A clear resemblance emerges
upon inspection, and the agreement becomes even more evident with the results
in Fig.~\ref{fig7}, where the \vmic\ velocities of the sample stars are directly  over-plotted 
on the corresponding  MTPF values derived from models that incorporate
both stellar rotation and turbulent pressure
\footnote{These models were computed by one of us (LG), following the methodology of
\citet{grassitelli15a} but including the effects of rotation at \vrot(init) = 300~\kms.}.
This  qualitative result was further confirmed by statistical tests which indicate that 
the empirical \vmic\ and the predicted MTPF are strongly correlated: 
$\rho\gtrsim$0.8 for both the high and low mass targets.

Although still preliminary, these findings suggest a strong physical link between
photospheric micro-turbulence and envelope convection, with the amplitude of turbulent
pressure fluctuations (TPF) providing a useful indicator. The broad \vmic\ maximum seen
in higher-mass OB stars appears to be tied to the peak TPF in the FeCZ
(log$\La$/$\La{_\odot}$$\gtrsim$3.4 --3.5). In contrast, AF(G)-type dwarfs, giants, and
supergiants (including classical Cepheids) exhibit a similarly pronounced peak that may
stem from significant TPF ($\gtrsim$10\%) in the hydrogen convection zone. This convection
zone is predicted to emerge from the ZAMS at log$\La$/$\La{_\odot}$$\approx$1.0
and extend up to the more luminous supergiants with 3.60$\lesssim$log\Teff$\lesssim$3.85 (see Fig.2
of \citealt{cantiello19} and preliminary results in Fig.~\ref{fig7}).

For main-sequence B stars (including $\beta$ Cephei variables and SPBs), it is noteworthy
that many exhibit non-negligible \vmic\ despite occupying a region where the turbulent
pressure contribution is predicted to be relatively small (below a few percent; see the
upper panel of Fig.~\ref{fig7}). One possible explanation is that a distinct, non-rotational
line-broadening mechanism unrelated to subsurface convection may be contributing to an
enhanced \vmic. Since most outliers lie in the part of the diagram where high-order
$g$-mode and low-order $p$-mode pulsations are expected (Sect.~\ref{vmic_puls};
Figs.~\ref{fig4} and \ref{figA1}), we propose that these oscillations could account for
the unusually large \vmic\ values observed in some MS B stars. In support of this idea,
\citet{SS17} found significant differences in line-profile shapes and variability between
MS B stars and OB supergiants, which they interpreted as empirical evidence for multiple
types of non-rotational line-broadening mechanisms operating in hot massive stars.

In summary, velocity fields generated by envelope convection zones \citep{cantiello09,
cantiello21, jiang15, grassitelli15a, grassitelli15b, schultz23} provide a
plausible explanation for both micro- and macro-turbulence. However, the specific mechanisms
linking envelope convection to the observed surface velocity fluctuations remain uncertain.
One possible scenario is that convective plumes retain their momentum as they enter thermally
diffusive, convectively stable layers \citep{jiang15}, resulting in surface velocities
proportional to those within the convective zone. Alternatively, pressure perturbations may
trigger high-order pulsations that propagate outward, thereby inducing velocity perturbations
at the stellar surface \citep{cantiello09, grassitelli15b}. To clarify exactly how
(sub)surface convection zones affect surface velocity fields, further radiation-
(magneto)hydrodynamics simulations of the outer envelopes of early-type stars are clearly
warranted \citep{jiang15, schultz20, schultz22, schultz23}.

\subsection {\it Micro-turbulence and the deficit of slow rotators among massive
OB stars}\label{vmic_vrot}

The near absence of slow rotators among massive O-type stars and early-B supergiants (SGs) has been recognized for decades, with \citet{CC77} among the first to highlight the issue. Although this phenomenon has often been attributed to the presence of non-rotational line broadening—commonly referred to as macro-turbulence — recent studies have shown that the problem persists even when the effects of \vmac\ are taken into account \citep{sundqvist13, markova14, SH14, SS17}. Current methodologies for determining projected rotational velocities (\vsini)—such as the Fourier Transform (FT), Goodness-of-Fit (GOF), and combined FT+GOF techniques—systematically neglect the role of micro-turbulence. Yet, theoretical models suggest that \vmic\ introduces an upper limit to \vsini, below which stars cannot be observed. This limit, referred to hereafter as \vsini~(upl), was introduced by \citet{SH14} and merits observational testing.

To constrain the proposed connection between \vmic\ and \vsini~(upl), we undertook the following approach. First, we compiled a large set of \vsini\ values from the literature for Galactic stars with various parameters, prioritizing those already included in our database (see below). Second, we analysed the behaviour of \vsini~(upl) as a function of \Teff, categorizing stars into the same four spectral luminosity subgroups previously used to examine \vmic. Third, we compared the trends of \vsini~(upl) and \vmic\ within each subgroup to identify potential correlations.

\begin{figure}[t]
\resizebox{\hsize}{!}{\includegraphics{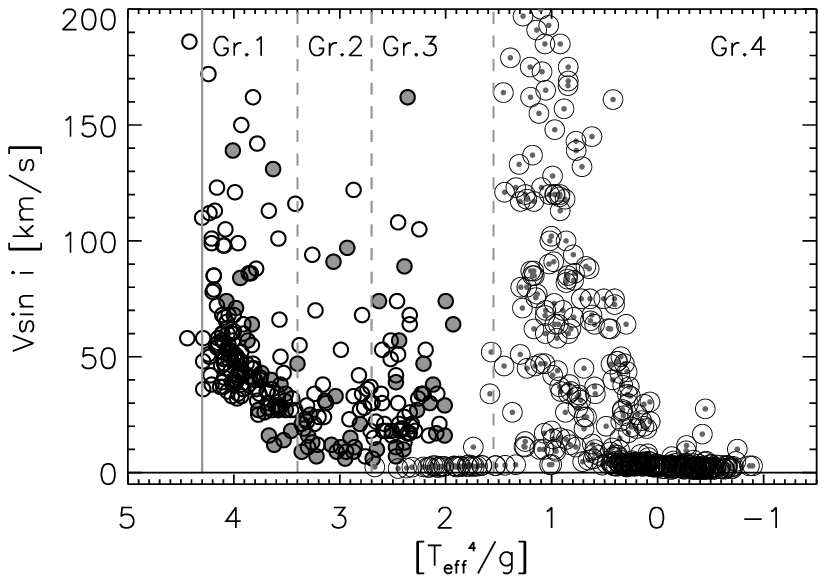}}
\caption{Projected rotational rates of Galactic stars of
various properties as a function of log$\La$/$\La{_\odot}$ (=[\Teff$^{4}$/$g$]) with vertical lines dividing the objects into four spectral luminosity subgroups as defined in Sect.~\ref{vmic_mass}. Different symbols are used to highlight the \vsini\ determinations which do (solid (DWs) and open 
(SGs) circles) and do not (small dots surrounded by a ring) account for the effect of macro-turbulent broadening (for more information see Sect.~\ref{vmic_vrot}).
}
\label{fig8}
\end{figure}

\begin{figure}[ht]
{\includegraphics[width=7.5cm,height=4.2cm]{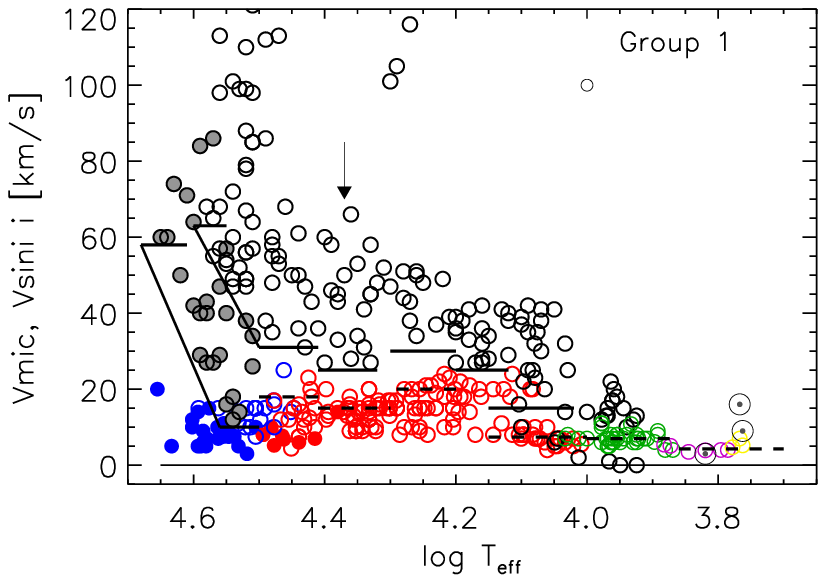}}\\
{\includegraphics[width=7.5cm,height=4.2cm]{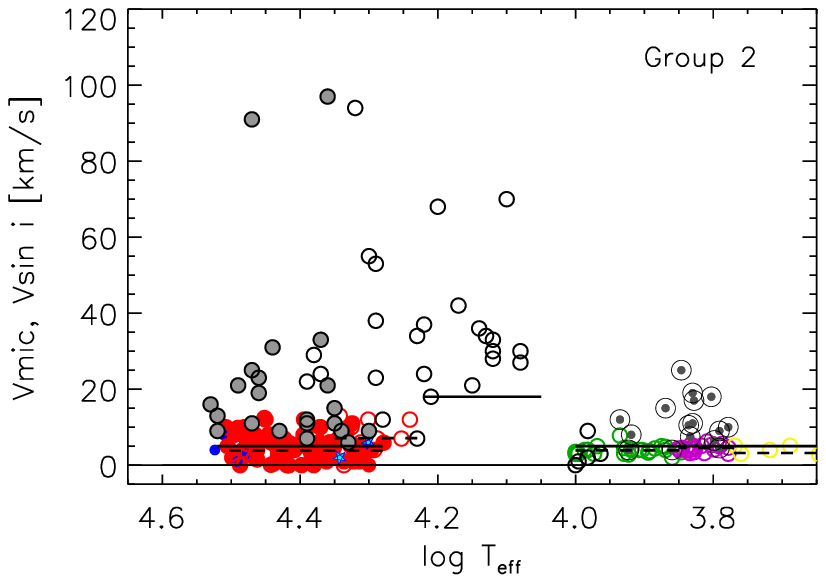}}\\
{\includegraphics[width=7.5cm,height=4.2cm]{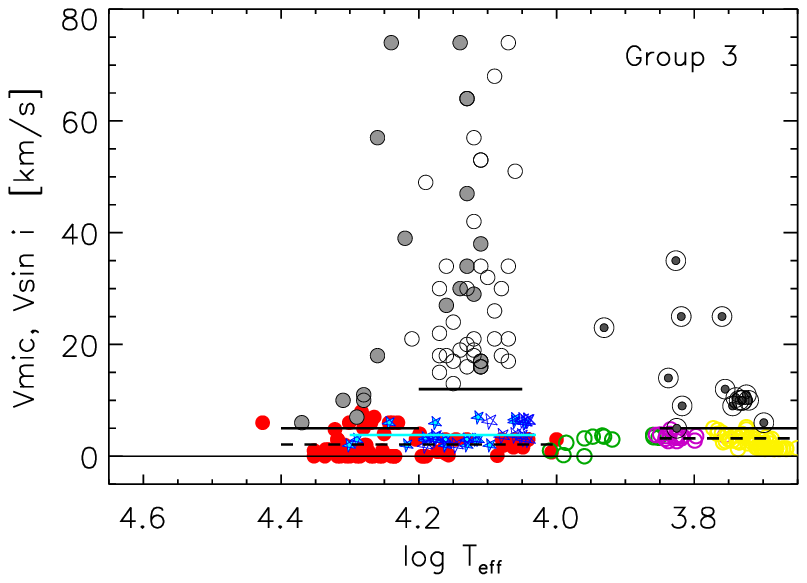}}\\
{\includegraphics[width=7.5cm,height=4.2cm]{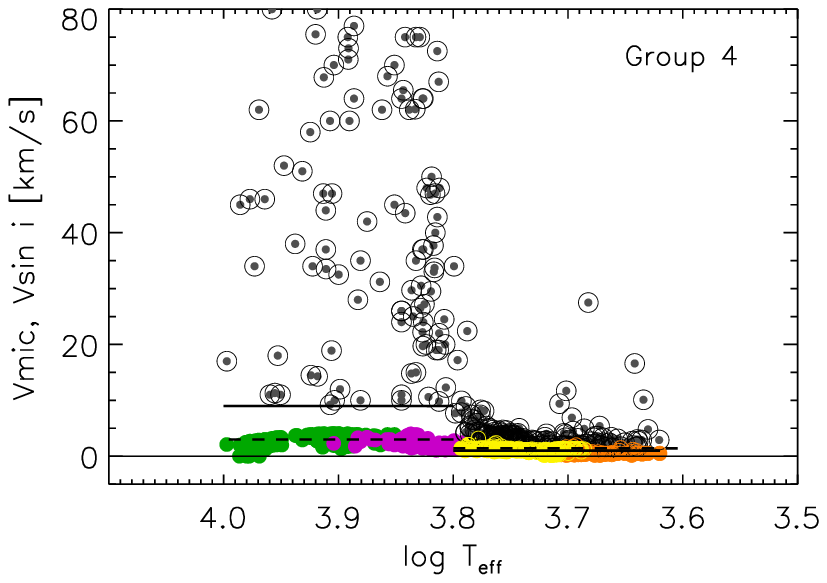}}
\caption{Micro-turbulent velocity of the sample stars separated by spectral
luminosity (Gr.1 to 4), and by SpT  and LC  (same symbols and
colours as in Fig.~\ref{fig3} with SPBs highlighted by a blue
five-pointed star). Overplotted in gray are the adopted \vsini\ 
(same symbols as in Fig.~\ref{fig8}). In each panel the horizontal solid and dashed
lines indicate, respectively, the mean value of the upper detectability
limit to \vsini\ averaged ''by eye", and the corresponding \vmean\ (for
more information see text).
}
\label{fig9}
\end{figure}

The key results of this analysis are presented in Figs.~\ref{fig8} and \ref{fig9}, and summarized in Table~\ref{vsini_vmic}. For all stars except those of OB spectral type, \vsini\ values (from \citealt{firnstein12, lyubimkov15, ivanyuk17, gebran10, luck18a, kahraman16, takeda08}) were derived in tandem with \vmic, enabling direct comparison. For OB stars, on the other hand, we used \vsini\ values published by \citet{SS17} (depicted as  solid (DWs)  and open (SGs) circles in both figures), resulting in an indirect comparison. It is important to note that for a non-negligible number of stars later than B-type (excluding luminous A-type SGs), the incorporated \vsini\ values do not account for macro-turbulent broadening. As a result, these values may be somewhat overestimated; such stars are marked as small dots surrounded by a ring in both figures (see \citealt{lyubimkov15, niemczura15} for further details).

While Fig.~\ref{fig8} reinforces earlier findings that the most massive O-type stars and B-type supergiants (Group 1) lack slow rotators, it also reveals that this phenomenon extends into the lower-mass B-star regime (Group 2). Across the entire high-mass domain ($\La/\La_{\odot}$$\gtrsim$3.3), \vsini(upl) decreases monotonically with decreasing spectroscopic luminosity, and thus with decreasing \Minit\ \citep[see][]{markova14}.

Fig.~\ref{fig9} further shows that in addition to \Minit, \vsini(upl) — or more precisely \vsinmean, defined here as a visual average of nearby stars with similar \vsini~(upl) (horizontal solid lines in each panel) — also depends on \Teff, with differing trends across the various mass subgroups. For Group~1 stars with log\Teff$\gtrsim$4.5 (typically of O spectral type), \vsinmean\ declines steeply toward cooler temperatures, with supergiants (open circles) consistently showing higher values than dwarfs (solid circles) of similar \Teff\ (see the two solid sloped lines in the top panel; see also \citealt{SH14}). In the remaining groups, a broad maximum in \vsinmean\ appears within the range 4.5$\lesssim$log\Teff$\lesssim$4.00 (corresponding to early- to late-type B supergiants), followed by a gradual decrease at cooler temperatures. For AF-type supergiants (log\Teff$\lesssim$4.0~dex), \vsinmean\ values tend to fall below 5~\kms, approaching zero on average.

For Group 2 and Group 3 stars, a notable deficit of slow rotators is observed among more evolved B-type stars (open circles, 4.2$\le$log\Teff$\le$4.0), whereas the remainder of the sample appears to be largely unaffected by this phenomenon.

Interestingly, even among low-mass AF(G)-type dwarfs (Gr. 4), our results reveal an absence of stars with low \vsini. While this finding could potentially be attributed to overestimated \vsini\ values due to neglected macro-turbulent broadening, we find this explanation unlikely. More massive and evolved AF-type giants and supergiants (Groups 1 to 3) —- which are more strongly affected by micro-turbulence (up to a factor of two; see Fig.\ref{fig3} and \citealt{gray84, gray87, firnstein12, ryabchikova15}) —- do not show any signs of a slow-rotator deficit.

Regarding the proposed relationship between \vsini~(upl) and \vmic — more specifically, between \vsinmean\ and \vmean\ (in each panel the solid and dashed horizontal lines, respectively) — our results (Table~\ref{vsini_vmic}) indicate that in stars strongly influenced by micro-turbulence (\vmean$\gtrsim$5\kms, Gr.~1 excluding O-type stars), the two quantities vary in tandem. \vsinmean\ tends to be approximately 1.5 to 2.0 times larger than \vmean, and even subtle features, such as local depressions, appear in both distributions (see vertical arrow). In stars less affected by micro-turbulence (Groups 2, 3, and 4), the relationship is non-uniform. Most targets have projected rotational velocities near zero (\vsinmean~$\lesssim$5~\kms), with more evolved MS B stars and AF-type dwarfs being exceptions, showing a marked deficit of slow rotators.

As for the sample of O-type stars, while limitations in our dataset may influence the outcome (see Sects.\ref{distr_vmic} and \ref{vmic_spt}), the top panel of Fig.\ref{fig9} suggests that their \vsinmean\ behaviour contrasts with that of their \vmic. Specifically, \vsinmean\ decreases with decreasing \Teff, while  \vmic\ tends to increase.

Since the \vsini\ and \vmic\ values used in this analysis were independently derived, these results can be interpreted as suggestive evidence that in cooler O- and B-type supergiants, the observed lack of slow rotators may arise from the neglect of micro-turbulent broadening in FT/FT+GOF methods—as predicted by FASTWIND line-profile simulations \citep{SH14}. However, for O-type stars near the zero age main sequence (ZAMS), more evolved MS B-stars, and possibly AF-type dwarfs additional mechanisms likely contribute to the elevated \vsinmean\ relative to \vmean.

\subsection {\it Micro-turbulence  and the upper detectability limit
to macro-turbulent broadening  established in the OB star regime}\label{vmic_vmac}

The absence of OB stars with very low macro-turbulent velocity was first reported by \citet{lefever10} and \citet{markova14}, and subsequently confirmed by \citet{SS17}. In all these studies, the effect of micro-turbulent broadening was neglected in the derivation of \vmac. However, line-profile simulations with the FW code suggest that this omission can result in an upper detectability threshold for \vmac, below which stars would not be observed. This threshold is referred to hereafter as \vmac~(upl) \citep{SH14}. Consequently, the absence of low-\vmac\ OB stars may plausibly be attributed to neglected micro-turbulent broadening, a hypothesis that warrants careful investigation.
\begin{table*}[h]
\caption{Micro-turbulence and the upper detectability limit to macro-turbulent
broadening for Galactic stars of various parameters separated by  spectral type
(OBAFG), luminosity class (dw  -- dwarfs; g -- giants; sg --supergiants) and
initial stellar mass: Group 1 -- \Minit$\gtrsim$15\Msun;
Group 2 -- 8\Msun$\lesssim$\Minit$\lesssim$15\Msun; Group 3 --
3\Msun$\lesssim$\Minit$\lesssim$8~\Msun; Group 4 -- 3\Msun$\lesssim$\Minit$\lesssim$1\Msun.
}\label{vmac_vmic}
\footnotesize
\tabcolsep0.9mm
\begin{tabular}{l|cccccccc|ccc|cc|ccc}
\hline
   \multicolumn{1}{l}{parameter} &
  \multicolumn{8}{c}{Group 1 } &
  \multicolumn{3}{c}{Group 2} &
  \multicolumn{2}{c}{Group 3} &
  \multicolumn{3}{c}{Group 4} \\
  \hline
  \multicolumn{1}{c}{} &
  \multicolumn{2}{c}{O dw} &
  \multicolumn{2}{c}{O g/sg} &
  \multicolumn{3}{c}{B sg} &
  \multicolumn{1}{c}{A }&
  \multicolumn{2}{c}{B} &
  \multicolumn{1}{c}{AF }&
  \multicolumn{2}{c}{B}&
 \multicolumn{1}{c}{AF}&
 \multicolumn{1}{c}{G}&
 \multicolumn{1}{c}{K}
 \\
	   &hot &cool &hot &cool & hot & inter   &cool &sg   &dw  &g/sg &g/sg       &dw  &g      &dw  &dw &dw\\
  \hline
  \hline
  \vmean   & ...  &8$^{a}$  & ...  &18$^{a}$ &18$^{a}$ &15/20$^{a}$ &7.4(1.8)  &7.0(1.8)  &3.8(3.3) & ... & 4.5/3.9 &2.1(2.2) &3.8(1.6) $^{b}$    &3.0(2.4) &1.0(0.4)&0.6(0.3)\\
  \vmacmean&85  &30   &85  &50   &50   &50           &30   &20   &15  &20  &11/8 &17&17 &$\sim$10$^{c}$& ... & ...\\
\hline
\end{tabular}
\\
\\
\small
{\bf Notes.}
\vmean  -- mean value of micro-turbulent velocity,  averaged over the corresponding
mass and temperature regime and its standard error (in brackets); \vmacmean -- a
''by eye"  average over nearby stars of similar upper detectability limit to \vmac;
''a" -- "by eye" estimate; ''b" -- an estimate corresponding to the sample SPBs;
''c" - an extrapolation of the Gray 1984 \vmac -- \Teff\ calibration towards hotter
temperatures.
\end{table*}

To investigate this issue further, and following the strategy outlined in the previous section, we searched the literature for accurate and reliable \vmac\ determinations for our sample stars, prioritizing those obtained in parallel with \vmic, \Teff, and \logg. As a result, we found suitable data for all stars except those of OB type: for massive AFG-type supergiants (Groups 1 and 2), we used values from \citet{firnstein12} and \citet{gray87}; for low-mass FGK-type dwarfs (Group 4), we adopted values from \citet{saar97}, \citet{doyle13}, and \citet{gray84}. For OB stars, we used the \vmac\ values published by \citet{SS17}, excluding any stars for which \vmac\ was reported as an upper limit.

To avoid potential systematic discrepancies caused by the use of different model line profiles \citep{markova14, SH14, TU17}, we considered only \vmac\ values derived using the radial-tangential model. Additionally, for AF-type stars (excluding the most luminous supergiants), individual \vmac\ measurements were not available. Instead, we relied on mean values across defined temperature and gravity ranges (denoted as \vmacmean) from \citet{gray87}, which we used as an alternative basis for our analysis.

\begin{figure}[ht]
{\includegraphics[width=7.4cm,height=4.2cm]{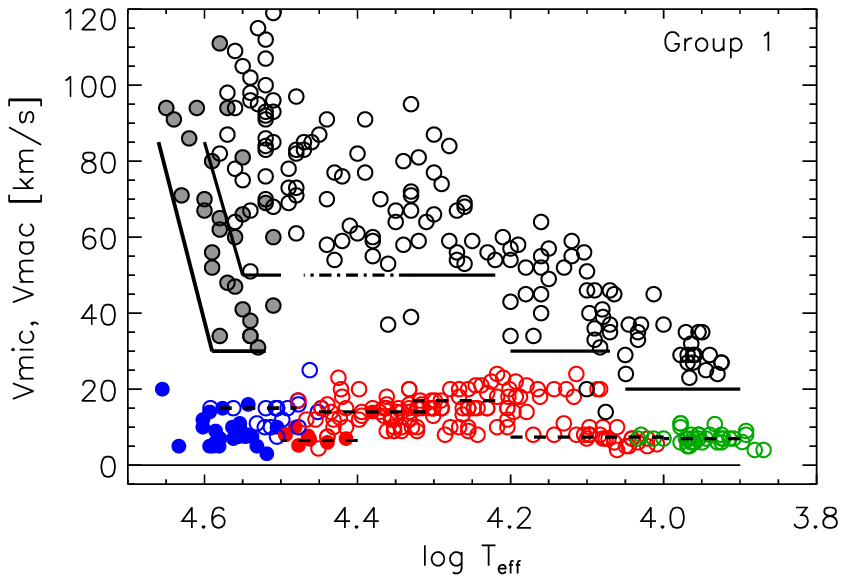}}\\
{\includegraphics[width=7.4cm,height=4.2cm]{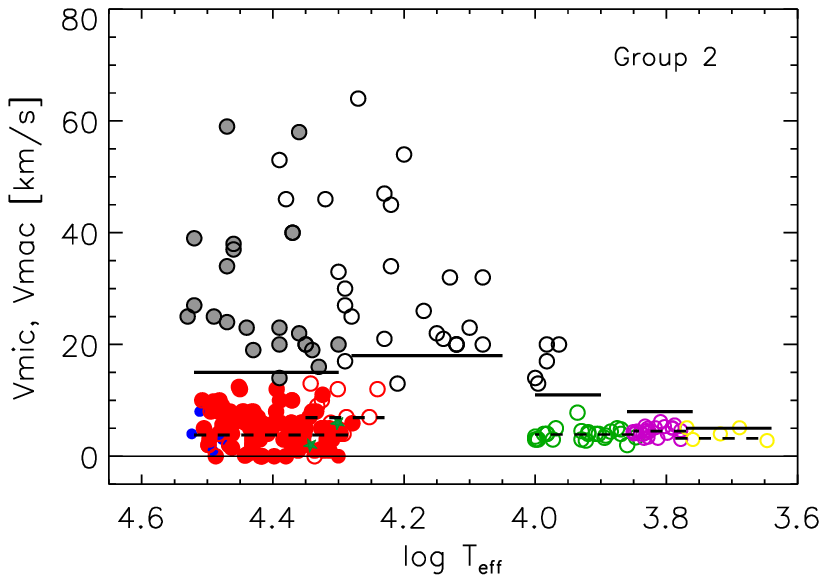}}\\
{\includegraphics[width=7.4cm,height=4.2cm]{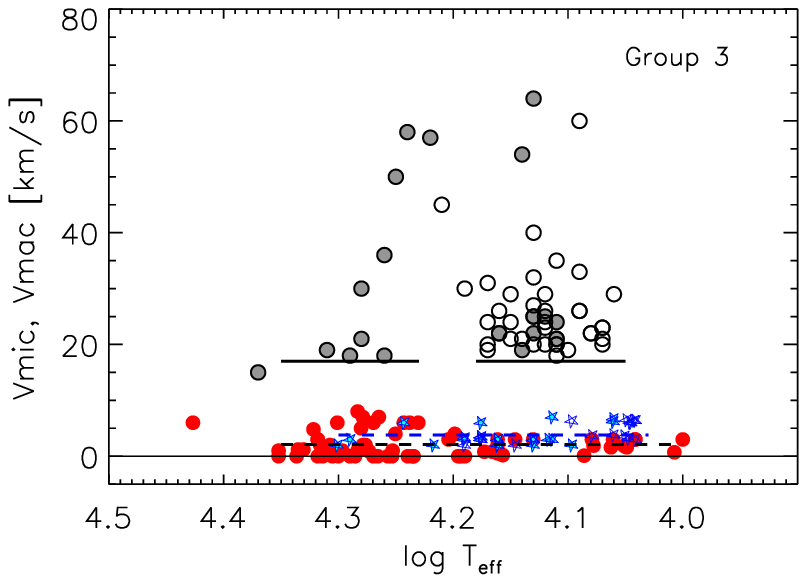}}\\
{\includegraphics[width=7.4cm,height=4.2cm]{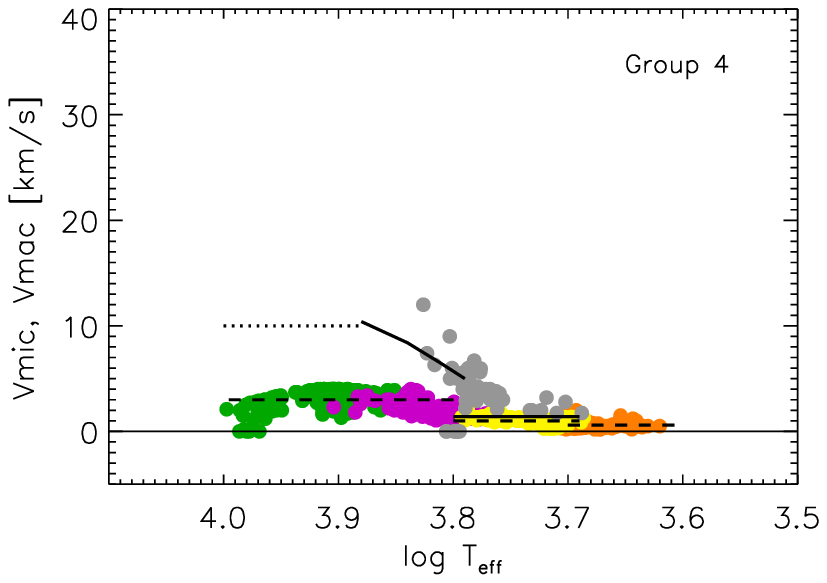}}\\
\caption{Micro-turbulent velocity  of the sample stars, separated by
spectral luminosity (Gr.1 to 4), and by SpT and LC as a function of
\Teff (same symbols and colours  as in Fig.~\ref{fig3}). Overplotted in gray
are the adopted \vmac\ data with solid and open circles representing
LC~V/IV and LC~III/I objects, respectively. The mean value of the upper
detectability limit to \vmac\ averaged ''by eye", and the corresponding
\vmean\ are also provided to guide the eye (horizontal solid and dashed
lines, respectively). In the bottom panel the solid curved line indicates
the \vmac--\Teff\  calibration provided by \citet{gray84}. For more
information see text.
}
\label{fig10}
\end{figure}

The main results of our \vmac--\vmic\ comparison, categorized by initial stellar mass (Groups 1 through 4) and  by spectral type (SpT) and luminosity class (LC), are illustrated in Fig.\ref{fig10} and summarized in Table\ref{vmac_vmic}. Key findings include:

\textbullet\ In addition to OB stars (Groups 1, 2, and 3; log\Teff~$\gtrsim$4.0), we observe a deficit of stars with very low macro-turbulent velocity (\vmacmean$<$5\kms) among massive AF-type giants and supergiants (Groups 1 and 2; 3.8$\lesssim$log\Teff~$\lesssim$~4.0), and possibly also among dwarfs of similar temperatures (see horizontal solid lines in each plot).

\textbullet\ For all stars except the hottest O-types  and AF-type dwarfs (discussed below), higher values of \vmacmean are generally correlated with higher \vmean. This correlation holds even for stars with relatively modest micro-turbulence (\vmean$\lesssim$5\kms), as evidenced by a strong Spearman correlation ($\rho$~=0.82, $\pi$=0.00; see Table\ref{vmac_vmic}).

\textbullet\ For the hottest O-type stars (log~\Teff~$\gtrsim$4.5; SpT$<$O9), \vmac(upl) tends to decrease with decreasing \Teff, with supergiants systematically showing higher values than dwarfs at the same temperature. This trend is depicted by two solid sloped lines in the top panel (see also Fig.9 in \citealt{SH14}). Notably, this behaviour contrasts with that of \vmic, which increases instead. 
While our sample of very massive O-stars near the ZAMS may be limited (see Sects.\ref{distr_vmic} and \ref{vmic_spt}), the main implication is that, in these stars, macro- and micro-turbulence may arise from different physical mechanisms -— an idea already suggested by \citet{AR15}.

\textbullet\ For AF-type dwarfs, the incomplete data coverage prevents firm observational constraints on \vmac\ properties. Nevertheless, based on the available measurements and the \vmac--\Teff\ calibration from \citet{gray84} (represented by the solid curved line in the bottom panel of Fig.\ref{fig10}), an increasing trend in \vmac\ toward higher temperatures is evident. A typical \vmacmean\ of approximately 10\kms\ at log~\Teff~$\gtrsim$3.8~dex (shown by the horizontal dotted line in the same plot) is consistent with the general behavior of \vmic.

Since the \vmic\ and \vmac\ values used in this analysis have been independently derived, the results outlined above suggest the following: First, a connection between small-scale and large-scale turbulent motions in stellar photospheres—whether direct (see, e.g., \citealt{mucciarelli11} and \citealt{husser13}) or indirect (through a common physical origin; see Sect.~\ref{vmic_conv})—appears likely and merits further empirical investigation. Second, although neglecting micro-turbulent broadening in the FT/FT+GOF method can contribute to the observed absence of stars with low \vmac, this factor alone is unlikely to fully account for the phenomenon, which is observed across stars of varying properties throughout the HR diagram.

\subsection{Micro-turbulence and the  mass problem in O stars}\label{mass_discrepancy}

The so-called mass problem —- that is, evolutionary masses (\Mevol) being larger than spectroscopic masses (\Mspec) -— is a well-documented phenomenon in single O-type stars as well as in members of spectroscopic and detached eclipsing binaries \citep{herrero92, heap06, mahy15, markova18, castro18, bestenleh20, mahy22, pavlovski23}. Despite considerable attention, the issue remains under debate \citep[see][]{serenelli21}, mainly due to the diversity of atmosphere models and evolutionary tracks employed to derive \Mspec\ and \Mevol, as well as the dependence on the choice of diagram (HR, sHR, or Kiel) used for comparison \citep{MP15, garland17, carolina17, markova18, castro21}.

While many studies point toward shortcomings in evolutionary models as the primary culprit —- such as missing or inadequately treated physics \citep{MP13, KPW17, pavlovski18, berlanas18, tkachenko20, johnston21, cole23} 
-— systematic uncertainties in spectroscopic parameters may also play a role \citep{sander15}. Notably, both the FW and CMFGEN codes neglect the contribution of micro-turbulent pressure in their hydrostatic or hydrodynamic balance equations. This omission likely leads to underestimated surface gravities and, therefore, underestimated \Mspec\ values \citep{SH98, MP15, mahy15, markova18, castro18, GT25} —- an effect warranting detailed investigation.

To explore the impact of neglected micro-turbulent pressure on derived gravities, we used our database to estimate the correction to \logg\ using a simple relation\footnote{
$\Delta$\logg = log $\left( 1 + \cfrac{ v_{\rm mic}^2\mu 60.5 }{T_{\rm eff}} \right) $
with $\mu\approx$ 0.6 (prvt. comm., Dr. J. Puls; see also \citealt{GT25})}
and estimated that for supergiants with log\Teff\ of 4.60~dex and 4.50~dex, an underestimation in this quantity by about 0.08 to 0.14~dex — corresponding roughly to \Mspec\ being about 0.28 to 0.37~dex lower than the actual value — can be expected. Analogous results for O-type dwarfs with similar temperatures indicate \logg\ and \Mspec\ values lower by about 0.02 to 0.01~dex and 0.14 to 0.10~dex, respectively
\footnote{As the presence of a positive gradient in atmospheric \vmic(r) has not been accounted for, these estimates should be considered as lower limits.}.

\paragraph{Stars with \Minit\ below 30–32\Msun}
For less massive O-stars (generally dwarfs and giants), there is ample empirical evidence suggesting that the mass problem is model-independent but may depend on the main photospheric characteristics of the stars. See, e.g., \citet{MP15} and \citet{markova18} for objects in the MW: $\Delta$log\Mspec\ of $\sim$0.09 to $\sim$0.12~dex; \citet{RA17} and \citet{carolina17} for objects in the LMC: $\Delta$log\Mspec\ of $\sim$0.08 and $\sim$0.10~dex, respectively; \citet{mahy22} for stars in the SMC: $\Delta$log\Mspec$=$0.10~dex (an estimate derived from their Fig.4); \citet{putkuri18} (see also \citealt{pavlovski23} and references therein): 0.16~dex$\lesssim\Delta$log\Mspec$\lesssim$0.09~dex
\footnote{Note that in all these works, optical observations combined with FW or CMFGEN modelling and rotating models from \citet{ekstroem12} and \citet{brott11} were consistently used.}.

Compared to these empirical findings, our estimates of $\Delta$log\Mspec(\vmic) are surprisingly close (and in the right direction!), suggesting that the mass discrepancy observed in less massive O-stars may, to a large extent, be mitigated if the FW and CMFGEN modeling properly accounts for the effect of microturbulent pressure.

\paragraph{Stars with \Minit$\gtrsim$40\Msun}
For more massive O-stars, the mass problem is strongly model-dependent and also sensitive to the main photospheric parameters, making it difficult to confirm or rule out definitively (see, e.g., \citealt{markova18, castro21, bouret21, berlanas18, bestenleh20} and references therein). For these reasons, no firm observational constraints can yet be placed on the effect of neglected micro-turbulent pressure on \Mspec. Nevertheless, using our database and results from Markova et al. as a reference
\footnote{We chose this work as a reference because: i) it is the first and so far only study where the effects of using different codes, evolutionary tracks, and types of diagrams have been systematically assessed in investigating the correspondence between \Mspec\ and \Mevol, and ii) all targets analyzed in it are included in our database, enabling a direct comparison.}
we suggest that if this effect is accounted for, it could either reduce the size of the discrepancy by $\sim$40\% and $\sim$70\% at log\Teff\ of 4.60~dex and 4.50 dex, respectively (Ekström et al. 2012 models with rotation) or increase it by $\sim$0.28~dex to $\sim$0.20~dex, the latter value corresponding to supergiants with log\Teff$\approx$4.48 (Brott et al. models with \vrot=300~\kms).

Although not perfect, these estimates may serve to place rough empirical constraints on any attempt to explain the lack of consistency between \Mspec\ and \Mevol\ in more massive O-stars when their properties are derived using the FW and CMFGEN codes.

\section{Summary and conclusions}\label{summary}

We have compiled an empirical database of \Teff, \logg, \vsini, \vmac, and \vmic\ determinations for more than 1800 presumably single stars with diverse properties in the MW. Both non-pulsating and pulsating objects are included, with the latter comprising $\sim$32\% of the total sample and encompassing six types of pulsators: $\beta$~Cephei, SPBs, $\gamma$~Doradus, $\delta$~Scuti, classical Cepheids, and Red Giants/Supergiants.

Based on these data, we study the behaviour of photospheric micro-turbulence as a function of \Teff\ and \logg\ within each SpT, LC, and pulsation type, and provide the first comprehensive and statistically significant overview of this phenomenon in Galactic stars. As a first application, we place observational constraints and evaluate several scenarios proposed to explain specific phenomena in hot, massive stars whose nature and origin remain poorly understood. The main results of our analysis are summarized below:\

\textbullet{Photospheric micro-turbulence is a universal spectral feature observed across all stars, regardless of their SpT, LC, or pulsational properties. Since the characteristics of this feature do not depend on the analysis method used, we conclude that the \vmic\ phenomenon is not an artifact of a specific modelling approach, but rather reflects a genuine physical process whose influence on photospheric line formation is not yet incorporated in current atmospheric models (Sections~\ref{vmic_spt} and \ref{vmic_puls}).}

\textbullet{The properties of photospheric micro-turbulence depend significantly on \Teff\ and \logg. However, the behaviour is too diverse to permit a deeper understanding when studied within a single SpT (Sect.\ref{vmic_spt} and Table\ref{stat_prop}).}

\textbullet{Preliminary results from a direct comparison between observed photospheric micro-turbulence in our sample and the amplitude of turbulent pressure fluctuations in models (accounting for stellar rotation with \vrot(init),$=$,300~\kms\ and turbulent pressure) strongly suggest that subsurface convection is the primary mechanism behind small-scale velocity fields in the photosphere. This idea, initially proposed by \citet{edmunds78} and \citet{cantiello09}, remains poorly understood in terms of its detailed mechanisms. Two plausible scenarios include: (i) high-order pulsations propagating outward and inducing surface velocity perturbations \citep{cantiello09, grassitelli15b}; and (ii) convective plumes retaining momentum as they penetrate thermally diffusive, convectively stable layers \citep{jiang15, schultz20, schultz22}. Further clarification will require dedicated radiation-(magneto)hydrodynamic simulations of early-type stellar envelopes.}

\textbullet{Using our database, we placed observational constraints on two specific phenomena observed in hot, massive stars whose origin is still unclear: the deficit of slow rotators and the absence of stars with very low \vmac\ velocities. Our results indicate that: (i) these phenomena are not limited to more massive OB stars, as previously thought, but also involve objects with markedly different characteristics, including more evolved LCIII B-stars and AF-type dwarfs; and (ii) neglecting micro-turbulent broadening in the FT and combined FT$+$GOF methods used to derive \vsini\ and \vmac\ — while potentially contributing, as suggested by \citet{SH14} — cannot solely account for the observed lack of stars with very low \vsini\ and \vmac\ in certain HR diagram regions (Sections\ref{vmic_vrot} and \ref{vmic_vmac}).}

\textbullet{Rough estimates of the effect of micro-turbulent pressure on surface equatorial gravities derived from observations strongly suggest that the mass discrepancy observed in O-stars with \Minit$\lesssim$30–32\Msun\ can be largely — if not entirely — resolved, provided that the microturbulent pressure term is properly included in 1D FW/CMFGEN modelling. This is consistent with the findings of \citet{GT25}, based on multidimensional O-star model atmospheres.}

\textbullet{Although previous studies have reported micro-turbulent velocities exceeding the sound speed in a few OB supergiants, our analysis provides the first strong empirical evidence that such small-scale supersonic motions are common among OB supergiants with log\Teff$\gtrsim$4.15~dex. As these velocity fields can drive key physical processes in massive stars
\footnote{e.g., shock formation affecting the energy balance in outer layers; shock-driven wind components complementing line-driven winds; small-scale density perturbations that may result in clumping, porosity, or altered ionization balance},
this finding highlights the need for more realistic theoretical simulations and evolutionary models that explicitly incorporate the presence of small-scale supersonic motions in the outer envelopes of early-type stars.}

\begin{acknowledgements}
NM thanks Dr. A. Herrero for useful comments and suggestions for improvements on the
manuscript. The IANAO (BAS) acknowledges financial support  from the Ministry
of Education and Science of Republic of Bulgaria trough the National Roadmap project.
The Center for Computational Astrophysics at the Flatiron Institute
is supported by the Simons Foundation.
\end{acknowledgements}

\begin{appendix}
\section{Micro-turbulence and stellar pulsations}\label{A}
\begin{figure*}[t]
\begin{center}
\resizebox{\hsize}{!}{\includegraphics{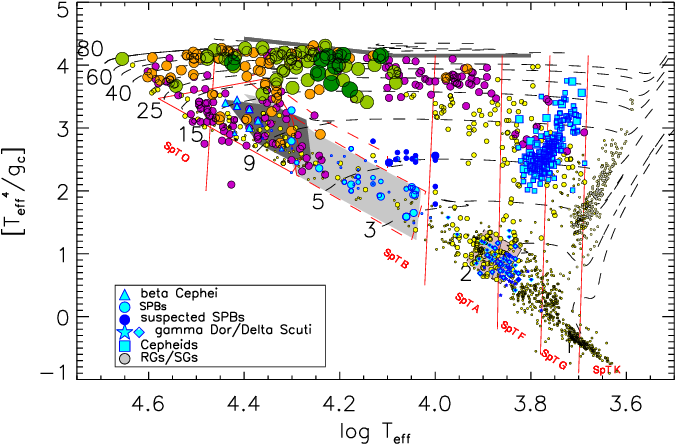}}
\caption{Same as Fig.~\ref{fig2} but with pulsating stars additionally highlighted
as indicated in the legend.  The theoretical instability strips for $\beta$
Cephei and SPBs from \citet{miglio07} (dark/light grey shaded  areas) and
from  \citet{moraveji16} (red solid/dashed line bordered rectangles),
and the instability domain for $\gamma$~Doradus and $\delta$~Scuty
variables  (from \citealt{RB01}, light pink shaded area) are also provided
for a direct comparison.
}
\label{figA1}
\end{center}
\end{figure*}

A total of 363 Galactic stars identified as radial pulsators (classical Cepheids) and non-radial pulsators (Slowly Pulsating B stars (SPBs), $\beta$~Cephei, $\gamma$~Doradus, and $\delta$~Scuti variables) have been included in our analysis. To this sample, we also added a large number of variable stars in the Red Giant/Supergiant (RG/SG) phase to cover the later stages of red-ward evolution on the HR diagram.

While the relatively small number of non-radial pulsators (non-RPs) — 160 in total — might seem limiting for our investigation, the lower panel of Fig.~\ref{fig1} shows that: i) the balance between radial and non-radial pulsators is nearly even, at 55\% versus 45\%, respectively; and ii) their distribution by oscillation type is reasonable, although the subsample of $\beta$~Cephei variables is somewhat under-represented. Furthermore, the positions of the non-RPs on the sHR diagram generally align well with theoretical predictions for their respective oscillation types, with most of the relevant parameter space being reasonably well covered (with the exception of $\beta$ Cephei stars; see Fig.\ref{figA1}). The same applies to the Cepheid instability strip and the Red Giant Branch, both of which are also well represented. These observations support the conclusion that our dataset enables a meaningful and realistic analysis of the \vmic\ properties of pulsating/variable stars and red giants.

The main results  of the \vmic\ analysis for the sample pulsators (grouped by oscillation type) and RGs are illustrated in Figures~\ref{fig4} and \ref{figA1}, and summarized in Table~\ref{stat_propP}. Note that for the non-RPs, we performed a comparison with non-pulsating stars of similar \Teff\ and \logg\ (hereafter referred to as "reference stars") to detect potential differences —- if any —- attributable to the respective oscillations.

\subsection{Non-radial pulsators}\label{nonP}
\subsubsection{Beta Cephei  and Slowly Pulsating B stars}

$\beta$ Cephei and SPBs are B-type pulsators which oscillate, respectively,
in coherent low-order pressure ($p$) \citep{SH05} and high-order gravity ($g$)
\citep{W91} modes excited by the $\kappa$-mechanism arising from the iron
opacity bump at \Teff $\approx$ 200~kK. % (however see also \citealt{mathis14}).
Both are MS stars with the former, on average, more massive and hotter than
the last: B0-B3 vs. B3-B9, and 8\Msun$\lesssim$\Minit$\lesssim$25\Msun vs.
3\Msun$\lesssim$\Minit$\lesssim$10\Msun.

\paragraph{$\beta$ Cephei}

There are 14 $\beta$ Cephei variables  in our sample  with five of them
recently suspected to be members of this subgroup \citep{burssens20}. While
the distribution of these stars is compatible with the instability strip
for  $p$ modes oscillations calculated by \citet{miglio07} and  by \citet{moraveji16}
\footnote{To account for recent results from \citet{bailey15}, theoretical
predictions based on two significantly different metal opacities -- from the
Opacity Project \citep{miglio07} and with Fe and Ni opacity increased by 75\%
each \citep{moraveji16} have been both considered in our study.}
(in Fig.~\ref{figA1}, dark gray  shaded area and the solid red line bordered
rectangle, respectively), the majority --if not all-- of them are located in
the high-mass portion of the predicted area (\Minit$\gtrsim$9\Msun), leaving
the low-mass region underpopulated.

From the top panels of Fig.~\ref{fig4} it appears that the range of \vmic\
for the sample of $\beta$ Cephei variables is qualitatively similar to
that of the reference stars, a result supported by our statistical
test (see corresponding data in Tables~\ref{stat_propP}).
Note that, due to the limited number of $\beta$~Cephei variables and their heterogeneous distribution across the corresponding \logg\ and \Teff\ intervals, these results should be interpreted with caution.

\paragraph{Slowly Pulsating B stars}
Sixteen bona fide SPBs and 37 candidates were initially included in our sample (all from \citealt{lefever10}). As shown in Fig.~\ref{figA1}, all but 13 of these stars (dark blue circles) lie within the instability strip for $g$-mode pulsations as computed by \citet{miglio07} and \citet{moraveji16} (light gray shaded area and the dashed red-bordered rectangle, respectively). Given that the photometric behavior of these outliers is comparable to that of the confirmed SPBs \citep{lefever10}, and that their \vmic\ properties are also similar (see corresponding data in Table\ref{stat_propP}), we retained these stars in the subsequent analysis to improve statistical robustness.

Interestingly, while the statistical properties of photospheric microturbulence in SPBs and their reference stars are quite similar, unlike the latter, none of the SPBs exhibit \vmic$\lesssim$2~\kms. Additionally, the temperature dependence of \vmic\ may differ between the two groups (see Tables~\ref{stat_propP}). If not attributable to observational selection effects or uncertainties in the derived stellar parameters, these findings may suggest that stellar pulsations caused by high-order $g$-mode oscillations influence the \vmic\ properties of SPBs. However, due to the absence of reference stars with \logg$<$3.5~dex, this possibility requires further investigation and independent verification.

\subsubsection{Delta Scuti and  Gamma Dor  variables}

Forty-four $\delta$~Scuti and 52 $\gamma$~Dor non-RPs are included in our analysis. On the sHR diagram, these stars are predominantly located within the theoretical $\delta$~Scuti/$\gamma$Dor instability region (light pink shaded area in Fig.~\ref{figA1}), although a non-negligible number appear somewhat cooler and less massive than expected. A detailed discussion of this issue can be found in \citealt{Uytt11, guzik21}.

From the third and fourth row panels of Fig.~\ref{fig4}, it appears that the sample $\delta$~Scuti and $\gamma$~Dor stars exhibit significant micro-turbulent broadening, with quite similar characteristics in both magnitude and in the dependence on \Teff\ and \logg. No evidence was found for notable differences in the \vmic\ properties of these non-radial pulsators compared to the corresponding reference stars (see data in Table\ref{stat_propP}).

\subsection{Classical Cepheids and Red Giants and Supergiants}\label{rp}

\paragraph{Classical Cepheids}
A total of 199 classical Cepheids are included in our sample. From Fig.~\ref{figA1}, it is evident that their positions on the sHR diagram fall entirely within the Cepheid instability strip (3\Msun$\lesssim$\Minit$\lesssim$20~\Msun and 3.70$\lesssim$\Teff$\lesssim$3.85). The majority of these stars ($\sim$83\% of the sample) are in the low-mass regime with \Minit$<$9~\Msun, that is not surprising, as low-mass stars evolve more slowly than their high-mass counterparts.

From the fifth row panels of Fig.~\ref{fig4}, we find that classical Cepheids exhibit significant micro-turbulent broadening, with an effect that appears to increase toward lower gravities and cooler temperatures (and hence higher \Minit). While this trend is qualitatively consistent with similar results for the non-variable stars of various SpT (see Sect.\ref{non_puls} and Fig.\ref{fig3}), it is not statistically confirmed: neither across the total sample nor within each of the two mass subgroups was any significant correlation between \vmic\ and \logg\ (or \Teff) found (see Table\ref{stat_propP} and the two horizontal solid lines in the fifth row panel on the left of Fig.~\ref{fig4}, which represent the \vmean\ of the high- and low-mass Cepheids).

Although it remains uncertain whether massive Cepheids are more strongly affected by micro-turbulence than their lower-mass counterparts, there are at least two possible factors that may influence the observed results:
i) observational selection effects, which tend to favour low-mass targets (as noted above); and
ii) significant scatter in \vmic\ at fixed \Teff\ and \logg\ — generally exceeding the 3$\sigma$ measurement error — likely caused by the combined use of “snapshot” and “phase-averaged” determinations of \Teff, \logg, and \vmic\ (see \citealt{proxauf18, luck18b} and references therein for more details).
More research with improved statistics, particularly for high-mass Cepheids, and predominantly phase-averaged \Teff and \logg values is needed to determine whether the properties of \vmic differ significantly between high- and low-mass Cepheids.

\paragraph{Red giants and supergiants}

A total of 222 stars identified as being in the red giant (RG) phase were initially included in our database. Based on their distribution on the sHR diagram (Fig.\ref{fig2}), all but seven lie below the 9\Msun\ evolutionary track and are thus classified as genuine RGs; the remainder are red supergiants (RSGs). Given the significant differences in the physics and evolutionary fate of red giants and supergiants (see, e.g., \citealt{GBP} and references therein), we made a clear distinction between these two subgroups to ensure robust analysis.

From the final row panels of Fig.~\ref{fig4}, two main features are immediately apparent:
First, the photospheric micro-turbulence of the RG and RSG samples (light and dark blue solid circles, respectively) is small but significant.
Second, although \vmic\ generally increases toward lower gravities and cooler temperatures, RSGs show systematically lower velocities — by about 0.5~\kms — than expected for their \logg\ values
\footnote{To estimate the expected \vmic\ for RSGs, the \logg--\vmic\ relationship for the RG sample was approximated by a linear regression of the form \vmic$=$1.94($\pm$0.06) $-$ 0.22($\pm$0.03)$\times$\logg.}
(see Table~\ref{stat_propP} and the red solid line in the bottom-left panel of Fig.~\ref{fig4}). These findings suggest that the \vmic\ properties of RSGs may differ from those of RGs —- a possibility that is not unexpected (see above) —- but given the limited number of RSGs in the sample, this conclusion remains tentative.

\section{Tables}\label{B}
\begin{table*}
\footnotesize
\begin{center}
\caption[]{Model atmosphere and radiative transfer codes used as a source
of \Teff, \logg and \vmic data for the sample stars}\label{B1}
\tabcolsep1.0mm
\begin{tabular}{llcllll}
\hline
\hline
 Name    &Type of        &Geometry    &Line blocking &Photosphere  &Wind   &References \\
         & model         &            & \& blanketing &            &      &\\
\hline
Model atmosphere codes\\
\hline
CMFGEN    &unified       &spherical       &yes     &from TLUSTY  &yes   &HM98\\
FASTWIND  &unified       &spherical       &approx. &yes          &yes   &P05\\
TLUSTY    &NLTE 	     &pp              &no      &yes          &no    &H98, LH07\\
ATLAS9    & LTE          &pp              &yes     &yes          &no    &K93\\
MARCS     & LTE          &spherical/pp    &yes     &yes          &no    &Gust08\\
\hline
Radiative transfer codes\\
\hline
DETAIL/SURFACE   &NLTE  & ...    & ...    &TLUSTY/ATLAS &no   & BG85\\
SYNSPEC   &NLTE         & ...   & ...    &TLUSTY       &no   &H98, LH07\\
SPECTRUM  &LTE          & ...    & ...    &ATLAS        &no   &GC94\\
WIDTH9    &LTE          & ...   & ...    &ATLAS9       &no   &K93\\
SYNTHE    &LTE          & ...   & ...    &ATLAS9       &no   &KA81\\
TGVIT     &LTE          & ...   & ...    &ATLAS9       &no    &T05\\
SSG       &LTE          & ...   & ...    &MARCS        &no    &B71, Gust08 \\
MOOG      &LTE          & ...    & ...    &MARCS        &no    &CK04\\
TURBOSPECTRUM &LTE      & ...   & ...   &MARCS        &no    &P12\\
\hline
\end{tabular}
\end{center}
{\bf Notes.} pp = plane-parallel;
HM98 = \citet{HM98}; P05 = \citet{P05}; BG85 = \citet{BG85}; H98 = \citet{H98};
LH07 = \citet{LH07}; KA81= \citet{kurucz81}, K93 = \citet{kurucz93}; Gust08 = \citet{gustaffson08};
GC94= \citet{GC94},B71 = \citet{bell71}, CK04=\citet{CK04}, P12= \citet{plez12}; T05=\citet{takeda05a}
\end{table*}
\begin{table*}
\begin{center}
\footnotesize
\caption[]{An overview of the sample stars with photometric \Teff\ and \logg\ determinations
and details regarding the corresponding analysis.
 }\label{photdat}
\tabcolsep0.50mm
\begin{tabular}{lrclllll}
\hline\hline
\multicolumn{1}{l}{Type of stars}
&\multicolumn{1}{l}{\# of }
&\multicolumn{1}{l}{Photom}
&\multicolumn{1}{l}{Type of}
&\multicolumn{1}{l}{Codes}
&\multicolumn{1}{l}{\vmic}
&\multicolumn{1}{l}{measurement error}
&\multicolumn{1}{l}{References}\\
\multicolumn{1}{l}{}
&\multicolumn{1}{c}{stars}
&\multicolumn{1}{c}{method}
&\multicolumn{1}{l}{models}
&\multicolumn{1}{l}{}
&\multicolumn{1}{l}{indicator}
&\multicolumn{1}{l}{$\Delta$\Teff/$\Delta$\logg/$\Delta$\vmic}
&\multicolumn{1}{c}{}\\
\hline
\hline
O9-B2 V        &22   &[M3a] &LTE/nLTE &ATLAS9/DETAL/SURFACE &O~II     &$\pm${\bf 1.20}/$\pm$0.10/$\pm$1.5  &DCB99, D01\\
B2-B3 IV/III   &4   &[M3a]  &LTE &ATLAS9/WIDTH9            &O~II      &{\bf $\pm$1.00/$\pm$0.20/$\pm$3.0}    &\citet{smartt02}\\
B2-A1 V/III    &45  &[M2]   &LTE &ATLAS9         	       & ....      &{\bf $\pm$0.40/$\pm$0.09/$\pm$1.0} &FM05\\
A5-G2 V/III/Ib &172 &[M2]   &LTE &ATLAS9$^{(a)}$/SPECTRUM  & ....     &{\bf $\pm$0.08/$\pm$0.10/$\pm$0.5}    &GGH01\\
A0-A8 V/III    &46  &[M1]   &LTE &ATLAS9$^{(b)}$       &O~I$^{d}$ &{\bf $\pm$0.30/$\pm$0.30/$\pm$0.3}    &\citet{takeda08}\\
AF V/IV        &67  &[M1]   &LTE &ATLAS9$^{(b)}$       &Fe~II/Mg~II&$\pm$0.12/$\pm$0.20/$\pm$1.0          &G08a, G08b, G10\\
AFG I/II       &63  &[M1]   &LTE &ATLAS9                   &Fe~II     &$\pm$0.12/$\pm${\bf 0.12}/$\pm$0.5  &L10, L15\\
FGK V          &158 &[M3b]  &LTE &MARCS/MOOG               &Fe~I      &{\bf $\pm$0.10/$\pm$0.10}/....                              &\citet{luck18a}\\
FGK V          &106 &[M3a]  &LTE &SAM12/WIOTA6$^{(c)}$     &Fe~I/II&   .... /$\pm${bf 0.12}/$\pm$0.2     &\citet{ivanyuk17}\\
RGs      &35  &[M3b]  &LTE &MARCS//MOOG              &Fe~I$^{d}$/Fe~II &{\bf $\pm$0.26/$\pm$0.12/ ....} &\citet{wang17}\\
\end{tabular}
\end{center}
\small
{\bf Notes.}
$\Delta$\Teff in kK; $\Delta$\logg\ in dex; $\Delta$\vmic\ in \kms. Bold faced numbers are
upper limits.\newline
(a) -- M. Lemke's distribution of ATLAS9.
(b) -- A modified version of WIDTH9 developed by \citet{takeda95a};
(c) -- Specific codes developed by \citet{pavlenko03};
(d) -- NLTE effects taken into account;
D01=\citet{daflon01}; DCB99=\citet{daflon99};
G08a=\citet{gebran08a}; G08b=\citet{gebran08b};
G10=\citet{gebran10}; GGH=\citet{gray01}; L10=\citet{lyubimkov10}; L15=\citet{lyubimkov15}; FM05=\citet{FM05}
\end{table*}

\begin{table*}
\footnotesize
\begin{center}
\caption[]{An overview of the sample stars with spectroscopically derived properties and  details
regarding the corresponding analysis.  }\label{spdat}
\tabcolsep0.5mm
\begin{tabular}{llllllll}
\hline
Type of stars      &\# of        &Type of  &Codes &Analysis    &\vmic\    &measurement error &References  \\
                   & stars$^{(a)}$ &models   &    &method  &indicators&$\Delta$\Teff/$\Delta$\logg/$\Delta$\vmic&\\
\hline
\hline
\\
High mass stars     \\
\hline
O9-B3 Ia            &18   &NLTE  &CMFGEN           &LPF  &all lines    &$\pm$1.00/$\pm${\bf 0.15}/ --- &CLW06\\
O7-O9.7 V/III/I     &14   &NLTE  &FASTWIND         &LPF  &He~I~6678$^{(b)}$ &$\pm$0.50/$\pm$0.10/$\pm$3.0 &MPL18\\
O4.5-O9.7 V/III/I   &14   &NLTE  &FASTWIND         &     &                  &$\pm$0.50/$\pm$0.10/$\pm$3.0 &\citet{holgado18}\\
O4-O9.5             &9    &NLTE  &FASTWIND         &EWs  &CNO          &$\pm$1.00/$\pm$0.10/$\pm${\bf 5.0} &carneiro19\\
O7-B0 V             &12   &NLTE  &FASTWIND         &EWs  &HHeSi       &$\pm$0.50/$\pm$0.10/$\pm$2.0     &Neg15\\
O9.7-B2 III/V       &8    &NLTE  &FASTWIND         &EWs  &Si~IV/III/II&$\pm$0.50/$\pm$0.10/$\pm$2.0 &\citet{berlanas18}\\
B0.5-B9 Ia/Iab/Ib   &11   &NLTE  &FASTWIND         &LPF  &Si~IV/III/II &$\pm$0.50/$\pm$0.10/$\pm$2.0 &MP08, markovaM08\\
B0-B9 Ia/Ib         &19   &NLTE  &FASTWIND         &LPF  &Si~III/II    &$\pm$0.50/$\pm$0.10/$\pm$5.0 &LPA07\\
B0-B9 V/III/I       &74   &NLTE  &FASTWIND         &LPF  &HHeSi       &$\pm${\bf 1.00}/$\pm$0.10/$\pm$4.0  &\citet{lefever10}\\
O9.5-B3 I/III/V    &56    &NLTE &TLUSTY/SYNSPEC       &EWs  &Si~III   &$\pm$1.00/$\pm${\bf 0.20}/$\pm$5.0  &T07, H07, H09\\
B3-B5 Ia/Iab        &6    &NLTE  &TLUSTY/DETAIL/SURFACE &EWs  &Si~III trpl&$\pm$1.00/$\pm$0.20/$\pm$5.0 &McE99\\
B0--B8 II/Ib/Iab/Ia &43   &NLTE &TLUSTY/SYNSPEC        &EWs  &Si~III trpl&$\pm$1.00/$\pm${\bf 0.20}/$\pm$1.0 &\citet{fraser10}\\
B8-A3 Ib/Iab/Ia     &35   &LTE/NLTE &ATLAS9/DETAIL/SURFACE&LPF &N~II/Fe~II &$\pm${\bf 0.20}$\pm$0.10/$\pm$1.0 &FP12\\
B1-B3 V/II/Ib       &8    &LTE/NLTE &ATLAS9/DETAIL/SURFACE&EWs &Si~III/II  &$\pm$1.00/$\pm${\bf 0.20}/ ---  &\citet{vranken00}\\
O9.5-B3 III/V       &15   &LTE/NLTE &ATLAS9/DETAIL/SURFACE &LPF &O~II     &$\pm$1.00/$\pm$0.15/$\pm${\bf 3.0} &Mor06, Mor08 \\
B0-B3 III/V         &30   &LTE/NLTE &ATLAS9/DETAIL/SURFACE &LPF &N~II/Fe~II &$\pm$0.20/$\pm$0.10/$\pm$1.0  &PNB08, NS11, NP12\\
O9-B3  V/III        &28   &LTE/NLTE &G87/H87           &EWs  &O~II  &{\bf $\pm$1.00/$\pm$0.10}/--- &K91, K94a, K94b\\
A0-F0 Ib/II         &22   &LTE  &ATLAS9/WIDTH9   &EWs  &Fe~I/II/Ti~II &{\bf $\pm$0.30/$\pm$0.10/$\pm$1.0}&\citet{venn95}\\
B8-F8 Ia/Iab/Ib     &20   &LTE  &ATLAS9/WIDTH9        &EWs &C~I/He~I$^{(c)}$&$\pm${\bf 0.65}/$\pm$0.30/$\pm${\bf 1.0} & TT95\\
\hline
\hline
\\
Low mass stars\\
\hline
AF III/V           &54  &LTE  &ATLAS9/SYNTHE      &LPF   &Fe~I/II   &{\bf $\pm$0.20/$\pm$0.20/$\pm$0.4} & N15, N17\\
A7-G0 V/IV         &96  &LTE  &ATLAS9/SYNTHE      &EWs   &Fe~I/II   &$\pm$0.20/$\pm$0.30/$\pm$0.4  &Kah16, Kah17\\
FGK V/IV           &67  &LTE  &ATLAS9/TGVIT       &EWs   &Fe~I/II   &$\pm$0.05/$\pm$0.10/$\pm$0.1 &\citet{takeda05b}\\
G6-G9 III          &58  &LTE  &ATLAS9/TGVIT       &EWs   &Fe~I/II   &$\pm$0.05/$\pm$0.10/$\pm$0.1 &\citet{takeda05a}\\
B2-A3.5 III/V      &12  &LTE  &LLmodels/SynthV    &LPF   &all lines$^{(c)}$ &$\pm$0.50/$\pm$0.12/$\pm${\bf 4.0} &\citet{lehmann11}\\
F1-G5~II/Ib        & 7  &LTE  &MARCS/sneden73     &EWs   &Fe~I/II   &$\pm$0.20/$\pm$0.25/$\pm$0.5  &GFP97\\
\hline
\hline
\\
Cepheids           &52  &LTE  &MARCS/SSG          &LDR/EW  &Fe~I/II &$\pm$0.10/$\pm$0.15/$\pm$0.3  &\citet{luck18b}\\
Cepheids           &11  &LTE  &ATLAS9/MOOG        &LDRs/EW &Fe~I/II &$\pm$0.10/                    &\citet{proxauf18}\\
Cepheids           &136 &LTE  &ATLAS9/MOOG        &LDRs/EW &Fe~II     &$\pm$0.15/$\pm$0.10/$\pm$0.5  &\citet{andrievsky13}\\
RGs          &19  &LTE  &MARCS/TURBOSPECTRUM&EWs   &Fe~I/II   &{\bf $\pm$0.12/$\pm$0.4}/$\pm$0.4 &\citet{palacio16}\\
RGs               &51  &LTE  &ATLAS9/TGVIT       &EWs   &Fe~I/II   &$\pm$0.05/$\pm$0.10/$\pm$0.1 &\citet{takeda15}\\
RGs               &104 &LTE  &ATLAS9/MOOG        &LPF   &Fe~I/II   &$\pm$0.05/$\pm$0.10/$\pm${\bf 0.3}&\citet{cordero14}\\
RGs           &13  &LTE  &ATLAS9/MOOG        &EWs   &Fe~I/II   &$\pm$0.08/$\pm$0.20/$\pm$0.1  &\citet{thygesen15}\\
\hline
\end{tabular}
\end{center}
\small
{\bf Notes.}
$\Delta$\Teff\ in kK; $\Delta$\logg\ in dex; $\Delta$\vmic\ in \kms. Bold faced numbers are upper limits.\newline
EWs -- equivalent width measurements;  LPF -- line profile fitting; LDR -- line depth ratio;
(a) -- The numbers given in this column  may be lower than  the total number of stars
investigated in the corresponding work (for more information see Section~\ref{data}).
(b) -- In O and very early B-type stars, this spectral line is very sensitive to \vmic \citep{lyubimkov04, markova20}.
(c) -- NLTE effects taken into account. \newline
CLW06=\citet{crowther06}; MPL18=\citet{markova18}; MP08=\citet{MP08}; markova08=\citet{markova08}; LPA07=\citet{lefever07}; McE99={McE99}; FP12=\citet{firnstein12}; TT95=\citet{takeda95b}; neg15=\citet{negueruela15}; N15=\citet{niemczura15};
N17=\citet{niemczura17}; Kah16=\citet{kahraman16}; T07=\citet{trundle07}; H07=\citet{hunter07}; H09=\citet{hunter09}; Mor06=\citet{morel06}; Mor08=\citet{morel08}; PNB08=\citet{prz08}; NS11=\citet{NS11};
NP12=\citet{NP12}; K91=\citet{kilian91}; K94a=\citet{kilian94a}; K94b=\citet{kilian94b};
GFP97=\citet{giridhar97}
\end{table*}
\end{appendix}

\begin{thebibliography}{}
\bibitem[Aerts et al.(2014)]{aerts14}
Aerts, C., Sim\'on-D\'iaz, S., Groot, P. J. and  Degroote, P., 2014, A\&A, 569, 118
\bibitem[Aerts \& Rogers(2015)]{AR15}
Aerts, C.,  and Rogers, T. M., 2015, ApJL, 806, 33
\bibitem[Andrievsky et al.(2013)]{andrievsky13}
Andrievsky, S. M.,  Lepine,J. R. D.,  Korotin, S. A. et al. 2013 MNRAS, 428, 3252
\bibitem[Asplund et al.(2000a)]{asp20a}
Asplund, M., Nordlund, A., Trampedach, R.  et al., 2020a, A\&A, 359,729
\bibitem[Asplund et al. (2000b)]{asp20b}
Asplund, M., Nordlund, A., Trampedach, R. et al. 2020b, A\&A, 359, 743
\bibitem[Bailey et al.(2015)]{bailey15}
Bailey, J. E., Nagayama, T., Loisel, G.P. et al. 2015, Nature, 517, 56
\bibitem[Bell(1971)]{bell71}
Bell, R. A. 1971, MNRAS, 154, 343
\bibitem[Berlanas et al.(2018)]{berlanas18}
Berlanas, S.R., Herrero, A., Comerón, A. et al. 2018 A\&A, 620, 56
\bibitem[Bestenlehen et al.(2020)]{bestenleh20}
Bestenlehner, J.M., Crowther, P.A., Caballero-Nieves, S.M. et al. 2020, MNRAS, 499, 1918
\bibitem[Bouret et al.(2005)]{bouret05}
Bouret, J.-C., Lanz, T. \&  Hillier, D.J. 2005, A\&A, 438, 301
\bibitem[Bouret et al.(2021)]{bouret21}
Bouret, J.-C., Martins, F., Hillier,W.L.  et al. 2021, A\&A, 647, 137
\bibitem[Brott et al.(2011)]{brott11}
Brott, I., de Mink, S. E., Cantiello, M., et al. 2011, A\&A, 530, A115
\bibitem[Burssens et al.(2020)]{burssens20}
Burssens, S., S. Sim\'on-D\'iaz, S., Bowman, S.,D. et al. 2020, A\&A, 639, 81
\bibitem[Butler \& Giddings(1985)]{BG85}
Butler, K. \& Giddings, J.R. 1985, Newsl. Anal. Astron. Spectra, No. 9
\bibitem[Cantiello et al.(2009)]{cantiello09}
Cantiello, M., Langer, N., Brott, I., et al. 2009, A\&A, 499, 279
\bibitem[Cantiello and Braithwaite(2019)]{cantiello19}
Cantiello, M., Braithwaite  2019, ApJ, 883, 106
\bibitem[Cantiello et al.(2021)]{cantiello21}
Cantiello, M., Lecoanet, D., Jermyn, A.S. et al. 2021, ApJ...915..112C
\bibitem[Carneiro et al.(2019)]{carneiro19}
Carneiro, L.P., Puls, J.,  Hoffmann, T.L. et al. 2019, A\&A, 623, 3
\bibitem[Castelli \& Kurucz(2004)]{CK04}
Castelli, F. \& Kurucz, R. L. 2004, A\&A, 419, 725
\bibitem[Castro et al.(2014)]{castro14}
Castro, N., Fossati, L., Langer, N. et al. 2014 A\&A 570, 13
\bibitem[Castro et al.(2018)]{castro18}
Castro, N., Oey, M. S., Fossati, L., \& Langer, N. 2018, ApJ, 868, 57
\bibitem[Castro et al.(2021)]{castro21}
Castro, N., Crowther, P.A. , Evans, C.J. et al. 2021, A\&A, 645, 68
\bibitem[Cole et al.(2023)]{cole23}
Cole, J., Mathias, M., Evan H. et al., 2023, ApJ (arXiv:2312.08315e)
\bibitem[Conti and Ebbets(1977)]{CC77}
Conti, P. S., \& Ebbets, D. 1977, ApJ, 213, 438
\bibitem[Cordero et al(2014)]{cordero14}
Cordero, M.J.,  Pilachowski, C.A., Johnson, C.I. et al. 2014, ApJ, 780, 94
\bibitem[Crowther, Lennon and Walborn(2006)]{crowther06}
Crowther, P. A., Lennon, D. J. \& Walborn, N.R. 2006 A\&A 446, 279
\bibitem[Daflon, Cunha \& Becker(1999)]{daflon99}
Daflon, S., Cunha, K. \&  Becker, S. 1999 ApJ, 522, 950
\bibitem[Daflon et al.(2001)]{daflon01}
Daflon, S., Cunha, K. \&  Becker, S. et al. 2001 ApJ, 552, 309
\bibitem[Doyle et al.(2013)]{doyle13}
Doyle, A. P., Davies, G. R., Smalley, B. et al. 2013, MNRAS, 428, 3164
\bibitem[Dufton et al.(2005)]{dufton05}
Dufton, P.L., Ryans, R.S.I., Trundle, C. et al. 2005, A\&A, 434, 1125
\bibitem[Edmunds(1978)]{edmunds78}
Edmunds, M. G. 1978, A\&A, 64, 103
\bibitem[Ekstr{\"o}m et al.(2012)]{ekstroem12}
Ekstr{\"o}m, S., Georgy, C., Eggenberger, P., et~al. 2012, A\&A, 537, 146
\bibitem[Fitzpatrick \& Massa(2005)]{FM05}
Fitzpatrick, E.L. \& Massa, D. 2005 AJ 129, 1642
\bibitem[Garland et a.(2017)]{garland17}
Garland, R., Dufton, P.L., Evans, C.J. et al, 2017, A\&A, 603, 91
\bibitem[Gebran \& Monier(2008)]{gebran08a}
Gebran, M. \& Monier, R. 2008 A\&A, 483, 567
\bibitem[Gebran, Monier \& Richard(2008)]{gebran08b}
Gebran, M. \& Monier, R. 2008 A\&A, 479, 189
\bibitem[Gebran et al.(2010)]{gebran10}
Gebran, M., Vick, M., Monier, R. et al. 2010, A\&A, 523, 71
\bibitem[Gebran et al(2014)]{gebran14}
Gebran, M., Monier, R., Royer, F. et al. 2014, In conference proceedings
"Putting A Stars into Context: Evolution, Environment, and Related Stars",
Moscow, 2013
\bibitem[Giridhar, Ferro \& Parrao(1997)]{giridhar97}
Giridhar, S., Ferro, A. \& Parrao, L. 1997, PASP, 109, 1077
\bibitem[Godart et al.(2017)]{godart17}
Godart, M., Sim\'on-D\'iaz, S., Herrero, A. et al. 2017, A\&A, 597, 123
\bibitem[Goldberg, Bildsten \&  Paxton (2020)]{GBP}
Goldberg, J.A., Bildsten, L. \&  Paxton, B. ApJ, 891, No.1
\bibitem[Gonz\'alez-Tor\'a et al.(2025)]{GT25}
Gonz\'alez-Tor\'a, G.; Sander, A. A. C.; Sundqvist, J. O. et al. 2025, A\&A, 694, 269
\bibitem[Grassitelli et al.(2015a)] {grassitelli15a}
Grassitelli, L., Fossati, L., Sim\'on-D\'iaz, S. et al. 2015, ApJ, 808, L31
\bibitem[Grassitelli et al.(2015b)] {grassitelli15b}
Grassitelli, L., Fossati, L., Langer, N. et al. 2015, A\&A, 584 , L2
\bibitem[Gray(1984)]{gray84}
Gray, R.O., 1984, ApJ, 281,719
\bibitem[Gray \& Toner(1987)]{gray87}
Gray, R.O. \& Toner, C.J. 1987, ApJ, 322,360
\bibitem[Gray\& Corbally(1994)]{GC94}
Gray, R.O., \& Corbally, C.J. 1994, ApJ, V107, 2
\bibitem[Gray, Graham \& Hoyt(2001)]{gray01}
Gray, R.O., Graham, P.W. \& Hoyt, S.R. 2001 AJ, 121, 2159
\bibitem[Gustafsson et al(2008)]{gustaffson08}
Gustafsson, B., Edvardsson, B., Eriksson, K et al. 2008, A\&A 486, 951
\bibitem[Guzik(2021)]{guzik21}
Guzik, J.A., Front. Astron. Space Sci., 2021, Sec. Stellar and Solar Physics, V8
\bibitem[Firnstein \& Przybilla(2012)]{firnstein12}
Firnstein, M. \& Przybilla, N. 2012 A\&A543,80
\bibitem[Fraser et al(2010)]{fraser10}
Fraser, M., Dufton, P.L., Hunter, I. et al 2010 MNRAS 404,1306
\bibitem[Haucke et al(2018)]{haucke18}
Haucke, M., Cidale, L. S., Venero, R. O. J. et al. 2018,A\&A, 614, 91
\bibitem[Heap et al.(2006)]{heap06}
Heap, S.R., Lanz, T. \& Hubeny, I. 2006, ApJ, 638, 409
\bibitem[Heiter \& Eriksson(2006)]{HE06}
Heiter, U.  \&  Eriksson, K. 2006, A\&A 452, 1039
\bibitem[Herrero et al.(1992)]{herrero92}
Herrero, A., Kudritzki, R.-P., Vilchez, J. M., et al.  1992, A\&A, 261, 209
\bibitem[Hillier \& Muller(1989)]{HM98}
Hillier, D.J. \& Muller, D.L. 1998, ApJ, 496, 407
\bibitem[Holgado et al.(2018)]{holgado18}
Holgado, G., Sim\'on-D\'iaz, S., Barb\'a, R.H. et al. 2018, A\&A, 613, 65
\bibitem[Holgado et al.(2020)]{holgado20}
Holgado, G., Sim\'on-D\'iaz, S., Haemmerl\'e, L. et al 2020, A\&A, 638, 157
\bibitem[Humphreys \& Davidson(1979)]{HD79}
Humphreys, R. \& Davidson, K. 1979, ApJ, 232,409
\bibitem[Hunter et al(2007)]{hunter07}
Hunter, I., Dufton, P.L., Smartt, S.J. et al. 2007, A\&A, 466, 277
\bibitem[Hunter et al.(2009)]{hunter09}
Hunter, I.,  Brott, I.,  Langer, N. et al. 2009 A\&A 496, 841
\bibitem[Hubeny(1998)]{H98}
Hubeny, I. 1998, ASPC, 138, 139
\bibitem[Husser et al.(2013)]{husser13}
Husser, T.-O., Wende von Berg, S., Dreizler, S. et al. 2013, A\&A 553, 6
\bibitem[Jiang et al. (2015)]{jiang15}
Jiang, Yan-Fey, Cantiello, M., Bilstend, L. et al.2015 ApJ 813, 74
\bibitem[Johnston(2021)]{johnston21}
Johnston, C. 2021A\&A, 655, 29
\bibitem[Jofre et al(2019)]{jofre19}
Jofre, P., Heiter, U. \& Soubiran, C. 2019 ARA\&A, 57, 571
\bibitem[Ivanyuk et al.(2017)]{ivanyuk17}
Ivanyuk, O.M., Jenkins, J.S., Pavlenko1, Ya.V. et al. 2017, MNRAS, 468, 4151
\bibitem[Kahraman et al.(2016)]{kahraman16}
Kahraman, A.F., Niemczura, E., De Cat, P. et al. 2016, MNRAS, 458, 2307
\bibitem[Keszthelyi, Puls \& Wade(2017)]{KPW17}
Keszthelyi, Z., Puls, J. \& Wade, Gr. 2017, A\&A, 598, 4
\bibitem[Kilian et al(1991)]{kilian91}
Kilian, J., Becker, S. R., Gehren, T. \& Nissen, P. E. 1991, A\&A, 244, 419
\bibitem[Kilian(1994)]{kilian94a}
Kilian, J. 1994, A\&A, 282, 286
\bibitem[Kilian, Montenbruck, Nissen(1994)]{kilian94b}
Kilian, J., Montenbruck, O., Nissen, P. E. 1994, A\&A, 284, 437
\bibitem[Kudritzki (1992)]{kud92}
Kudritzki, R.-P., 1992, A\&A, 266, 395
\bibitem[Kurucz(1993)]{kurucz93}
Kurucz, R. L. 1993, ATLAS9 Stellar Atmosphere Programs and
grid, CD-ROM No. 13 (Cambridge, MA: Smithsonian Astrophysical Observatory)
\bibitem[Kurucz \& Avrett(1981)]{kurucz81}
Kurucz, R.L. \& Avrett, E.H. 1981, SAOSR, 391
\bibitem[Langer \& Kudritztki(2014)]{LK14}
Langer, N. \& Kudritzki, R.-P. 2014, A\&A, 564, A52
\bibitem[Lanz \& Hubeny(2007)]{LH07}
Lanz, T. \& Hubeny, I. 2007, ApJS, 169, 83
\bibitem[Lehmann et al.(2011)]{lehmann11}
Lehmann, H., Tkachenko, A., Semaan, T. et al. 2011, A\&A 526, 124
\bibitem[Lefever, Puls, Aerts(2007)]{lefever07}
Lefever, C., Puls, J., Aerts, C. 2007, A\&A, 463, 1093
\bibitem[Lefever et al.(2010)]{lefever10}
Lefever, C., Puls, J., Morel, T. 2010, A\&A, 515, 74
\bibitem[Lyubimkov, Rostopchin, Lambert(2004)]{lyubimkov04}
Lyubimkov, L., Rostopchin, S. I., Lambert, D. L. 2004, MNRAS, 351, 745
\bibitem[Lyubimkov et al(2010)]{lyubimkov10}
Lyubimkov, L., Lambert, D.L., Rostopchin, S.I. et al. 2010, MNRAS, 402, 1369
\bibitem[Lyubimkov et al.(2015)]{lyubimkov15}
Lyubimkov, L.S., Lambert, D.L., Korotin, S.A. et al. 2015, MNRAS, 446,3447
\bibitem[Luck(2018a)]{luck18a}
Luck, R.E. 2018, AJ, 155, 111
\bibitem[Luck(2018b)]{luck18b}
Luck, R.E. 2018, AJ, 156, 171
\bibitem[Mahy et al.(2015)]{mahy15}
Mahy, L., Rauw, G., De Becker, M. et al. 2015, A\&A, 577, 23
\bibitem[Mahy et al.(2022)]{mahy22}
Mahy, L., Sana, H., Shenar, T. et al. 2022, A\&A, 664, A159
\bibitem[Markova \& Puls(2008)]{MP08}
Markova, N. \& Puls, J. 2008, A\&A, 478, 823
\bibitem[Markova \& Puls(2015)]{MP15}
Markova, N. \& Puls, J. 2015, IAUS, 307, 117
\bibitem[Markova, Puls \& Langer(2018)]{markova18}
Markova, N., Puls, J., Langer, N. 2018, A\&A, 613, 12
\bibitem[Markova et al.(2008)]{markova08}
Markova, N., Prinja, R.K. \& Markov, H. et al., 2008, A\&A, 487, 211
\bibitem[Markova et al.(2014)]{markova14}
Markova, N., Puls, J.,  Sim\'on-D\'iaz, S. et al. 2014, A\&A, 562, 37
\bibitem[Markova et al.(2020)]{markova20}  %
Markova, N., Puls, J., Dufton, P. et al. 2020, A\&A, 634, 12
\bibitem[Martins \& Palacios(2013)]{MP13}
Martins, F. \& Palacios, A. 2013, A\&A, 560, 16
\bibitem[Martins, Schaerer \& Hillier(2005a)]{martins05a}
Martins, F., Schaerer, D., Hillier, D.J. 2005a, A\&A, 436, 1049
\bibitem[Martins et al.(2005b)]{martins05b}
Martins, F., Schaerer, D., Hillier, D.J. et al. 2005b, A\&A, 441, 735
\bibitem[Massey et al.(2013)]{massey13}
Massey, Ph., Neugent, K. F., Hillier, D. J., \& Puls, J. 2013, ApJ, 768, 6
\bibitem[Massey, Neugent, Smart(2016)]{massey16}
Massey, Ph., Neugent, K. \& Smart, B., 2016, ApJ, 152, 62
\bibitem[McEvoy et al.(2015)]{Mcevoy15}
McEvoy, C. M.,  Dufton, P. L.,  Evans, C. J. et al. 2015, A\&A, 575, 70
\bibitem[McErlean, Lennon, Dufton(1999)]{McE99}
McErlean, N.D., Lennon, D.J., Dufton, P.L. 1999, A\&A, 349, 553
\bibitem[Miglio, Montalb\'an, Dupret(2007)]{miglio07}
Miglio, A., Montalb\'an \& Dupret, M.-A. 2007, MNRAS, 375, L21
\bibitem[Moravveji(2016)]{moraveji16}
Moravveji, E. 2016, MNRAS, 455, 67
\bibitem[Morel etal.(2006)]{morel06}
Morel, T., Butler, K., Aerts, C. et al. 2006, A\&A, 457, 651
\bibitem[Morel, Hubrig, Briquet(2008)]{morel08}
Morel, T., Hubrig, S., Briquet, M. 2008 A\&A 481, 453
\bibitem[Mucciarelli et al.(2011)]{mucciarelli11}
Mucciarelli, A., 2011, A\&A, 528, 44
\bibitem[Negueruela et al(2015)]{negueruela15}
Negueruela, I., Simon-Diaz, S., Lorenco, J et al. 2015, A\&A 584, 77
\bibitem[Niemczura et al.(2015)]{niemczura15}
Niemczura, E.,  Murphy, S.J., Smalley, B., et al. 2015, MNRAS, 450, 2764
\bibitem[Niemczura et al.(2017)]{niemczura17}
Niemczura, E., Polinska, M.,  Murphy, S.J.  et al. 2017, MNRAS, 470,2870
\bibitem[Nieva(2013)]{nieva13}
Nieva, M.-F. 2013, A\&A 550, 26
\bibitem[Nieva \& Przybilla(2007)]{NP07}
Nieva, M.F. \& Przybilla, N. 2007, A\&A 467, 295
\bibitem[Nieva \& Simon-Diaz(2011)]{NS11}
Nieva, M.-F. \& Simon-Diaz, S., 2011, A\&A, 532, 2
\bibitem[Nieva \& Przybilla(2012)]{NP12}
Nieva, M.F. \& Przybilla, N. 2012, A\&A 539, 145
\bibitem[Nieva \& Przybilla(2014)]{nieva14}
Nieva, M.F. \& Przybilla, N. 2014, A\&A 566, 7
\bibitem[Palacios et al.(2016)]{palacio16}
Palacios, A., Jasniewicz, G., Masseron, T. et al. 2016, A\&A, 587, 42
\bibitem[Pavlenko(2003)]{pavlenko03}
Pavlenko Ya.V., 2003, Astron. Rep., 47, 59
\bibitem[Pavlovski et al(2018)]{pavlovski18}
Pavlovski, K., Southworth, J. and Tamajo, E. 2018, MNRAS 481, 3129
\bibitem[Pavlovski et al(2023)]{pavlovski23}
Pavlovski, K., Southworth, J., Tkachenko, A. et al. 2023, A\&A, 671, 139
\bibitem[Proxauf et al(2018)]{proxauf18}
Proxauf, B., da Silva, R., Kovtyukh, V. V. et al. 2018, A\&A, 616, 82
\bibitem[Plez(2012)]{plez12}
Plez, B. 2012, https://ui.adsabs.harvard.edu/abs/2012ascl.soft05004P
\bibitem[Przybilla et al.(2001)]{prz01}
Przybilla, N.,  Butler, K., Becker, S.R. \& Kudritzki, R.-P., 2001, A\&A, 369, 1009
\bibitem[Przybilla et al.(2006)]{prz06}
Przybilla, N., Butler, K., Becker, S.R. et al. 2006, A\&A, 445, 1099
\bibitem[Przybilla, Nieva \& Butler(2008)]{prz08}
Przybilla, N., Nieva, M.F. \& Butler, K 2008, ApJ, 688, 10
\bibitem[Puls et al.(2005)]{P05}
Puls, J., Urbaneia, M.A., Venero, R. et al. 2005, A\&A, 435, 669
\bibitem[Putkuri et al.(2018)]{putkuri18}
Putkuri, C., Gamen, R.,Morrell, N.L. et al., 2018, A\&A 618,174
\bibitem[Ram\'irez-Agudelo et al.(2017)]{RA17}
Ram\'irez-Agudelo, O. H., Sana, H., de Koter, A., et al. 2017, A\&A, 600, 81
\bibitem[Rivero Gonz{\'a}lez et al.(2012)]{gonzalez12}
Rivero Gonz{\'a}lez, J.G., Puls, J., Najarro,  J.F. \& Brott, I. 2012, A\&A, 537, 79
\bibitem[Rodriguez \& Breger(2001)]{RB01}
Rodriguez, E. \& Breger, M. 2001, A\&A, 366, 178
\bibitem[Ryabchikova et al.(2015)]{ryabchikova15}
Ryabchikova, T., Piskunov, N. Pakhomov, Yu. et al. 2015, MNRAS, 456, 1221
\bibitem[Saar \& Osten(1997)]{saar97}
Saar, S. H. \& Osten, R. A., 1997, MNRAS, 284,803  %GK DWs
\bibitem[Sab\'in-Sanjuli\'an et al.(2017)]{carolina17}
Sab\'in-Sanjuli\'an, C., Sim\'on-D\'iaz, S., Herrero, A., et al. 2017, A\&A, 601, 79
\bibitem[Sander et al.(2015)]{sander15}
Sander, A., Shenar, T., Hainich, R. et al. 2015, A\&A 577, A13 (2015)
\bibitem[Sander(2017)]{sander17}
Sander, A.A.C. 2017, IAUS 329, 215
\bibitem[Schiller \& Przybilla (2008)]{SP08}
Schiller, F. \& Przybilla, N. 2008, A\&A, 479, 849
\bibitem[Searle et al.(2008)]{searle08}
Searle, S.C., Prinja, R.K., Massa, D. et al. 2008, A\&A, 481, 777
\bibitem[Serenelli et al.(2021)]{serenelli21}
Serenelli, A., Weiss, A.,  Aerts, C., Angelou, G.C. et al. 2021, A\&ARv, 29, 4
\bibitem[Sim\'on-D\'iaz \& Herrero(2014)]{SH14}
Sim\'on-D\'iaz, S. \& Herrero, A. 2014, A\&A 562, 135
\bibitem[Sim\'on-D\'iaz et al.(2017)]{SS17}
Sim\'on-D\'iaz, S., Godart, M., Castro, N. et al. 2017, A\&A 597, 23
\bibitem[Sim\'on-D\'iaz(2020)]{SS20}
Sim\'on-D\'iaz, S (2020)  A Modern Guide to Quantitative Spectroscopy of
Massive OB Stars. In: Reviews in Frontiers of Modern Astrophysics. Springer, Cham.
\bibitem[Schultz et al.(2020)]{schultz20}
Schultz, W., C., Bildsten, L., Jiang, Yan-Fey, 2020, ApJ 902, 67
 \bibitem[Schultz et al. (2022)]{schultz22}
Schultz, W., C.,  Bildsten, L., Jiang, Yan-Fey, 2022, ApJ 924, 11
\bibitem[Schultz et al. (2023)]{schultz23}
Schultz, W.C., Bildsten, L., Jiang, Yan-Fey, 2023 ApJ 951, 42
\bibitem[Smalley(2004)]{smalley04}
Smalley, B. 2004, The A star puzzle, in: Proceeding IAU Symposium
No. 224 %Zverko, J. %\'Zi\'z\'novsk\´y,  Adelman,S.J. \&  Weiss, W.W. (eds)
\bibitem[Smalley(2005)]{smalley05}
Smalley, B. 2005, Mem. S.A.It. Suppl. Vol. 8, 130
\bibitem[Smartt et al.(2002)]{smartt02}
Smartt, S.J., Venn, K.A., Dufton, P.L. et al. 2002, A\&A, 367, 86
\bibitem[Smith \& Howarth(1998)]{SH98}
Smith, K.C. \& Howarth, I.D. 1998, MNRAS, 299, 1146
\bibitem[Stankov \&  Handler(2005)]{SH05}
Stankov, A. and  Handler, G. 2005, ApJS, 158, 193
\bibitem[Sunqvist et al.(2013)]{sundqvist13}
Sundqvist, J. O., Sim\'on-D\'iaz, S., Puls, J. \& Markova, N.,2013, A\&A, 559, 10
\bibitem[Takeda(1995)]{takeda95a}
Takeda, Y. 1995, PASJ, 47, 287
\bibitem[Takeda \& Takada-Hidai(1995)]{takeda95b}
Takeda, Y. \& Takada-Hidai, M. 1995, PASJ, 47, 113
\bibitem[Takeda \& Ueno(2017)]{TU17}
Takeda, Y. \& Ueno, S. 2017 PASJ, 69, 46
\bibitem[Takeda et al.(2005a)]{takeda05a}
Takeda, Y.,  Sato, B., Kambe, E. et al. 2005a, PASJ, 57, 109
\bibitem[Takeda et al.(2005b)]{takeda05b}
Takeda, Y., Ohkubo, M., Sato, B. et al. 2005b, PASJ, 57, 27
\bibitem[Takeda et al.(2008)]{takeda08}
Takeda, Y., Han, I., Kang, D.-I.  et al. 2008, Journal of the KAS, 41, 83
%The Korean Astronomical Society, 41, 83
\bibitem[Takeda et al.(2015)]{takeda15}
Takeda, Y. \& Tajisu, A. 2015,  MNRAS, 450, 397
\bibitem[Townsend et al.(2007)]{townsend07}
Townsend, R., Owocki, S.P. \& Ud-Doula, A. 2007, MNRAS, 382, 139
\bibitem[Thygesen et al.(2014)]{thygesen15}
Thygesen, A.O., Sbordone, L., Andrievsky, S. et al. 2015, A\&A, 572, 108
\bibitem[Tkachenko et al. (2020)]{tkachenko20}
Tkachenko, N., Pavlovski, K.,  Johnston, C et al. 2020, A\&A, 637, 60 % MD in BS
\bibitem[Trundle et al.(2007)]{trundle07}
Trundle, C., Dufton, P.L., Hunter, I. et al. 2007, A\&A, 471, 625
\bibitem[Uytterhoeven et al.(2011))]{Uytt11}
Uytterhoeven, K., Moya, A., Grigahc\'ene, A. et al. 2011, A\&A, 534, A125
\bibitem[Venn(1995)]{venn95}
Venn, K. 1995, ApJSS, 99, 659
\bibitem[Verdugo, Talavera, G\'omez de Castro(1999)]{verdugo99}
Verdugo, E., Talavera, A.   \& G\'omez de Castro, A.I., 1999, A\&A, 346, 819
\bibitem[Vranken et al.(2000)]{vranken00}
Vranken, M., Lennon, D.J., Dufton, P.L., Lambert, D.L. 2000, A\&A, 358, 639
\bibitem[Wang et al.(2017)]{wang17}
Wang, Y., Primas, F., Charbonnel, C et al. 2017, A\&A, 607, 135
\bibitem[Waelkens(1991)]{W91}
Waelkens, C.  1991, A\&A 246, 453
\end{thebibliography}
\end{document}